\documentclass[aps,prd,eqsecnum,12pt,showpacs,nofootinbib]{revtex4-2}
\usepackage[utf8]{inputenc}
\usepackage{geometry}                		
\usepackage{graphicx}				
\usepackage{amssymb}
\usepackage{graphicx}
\usepackage{amssymb,amsmath}
\usepackage{amsfonts,mathrsfs}
\newcommand{\la}{\langle}
\newcommand{\ra}{\rangle}
\newcommand{\be}{\begin{equation}}
\newcommand{\ee}{\end{equation}}
\newcommand{\bea}{\begin{eqnarray}}
\newcommand{\eea}{\end{eqnarray}}
\newcommand{\bes}{\begin{subequations}}
\newcommand{\ees}{\end{subequations}}

\usepackage{amsmath}
\usepackage{ amssymb }
\usepackage{mathtools}
\usepackage{indentfirst}
\usepackage{graphicx}
\usepackage[
    colorlinks=true,
    linkcolor=blue
]{hyperref}
\usepackage{graphicx}
\usepackage[section]{placeins}
\usepackage{placeins}
\let\Oldsection\section
\renewcommand{\section}{\FloatBarrier\Oldsection}

\let\Oldsubsection\subsection
\renewcommand{\subsection}{\FloatBarrier\Oldsubsection}

\let\Oldsubsubsection\subsubsection
\renewcommand{\subsubsection}{\FloatBarrier\Oldsubsubsection}
\DeclarePairedDelimiter\abs{\lvert}{\rvert}
\DeclarePairedDelimiter\norm{\lVert}{\rVert}
\makeatletter
\let\oldabs\abs
\def\abs{\@ifstar{\oldabs}{\oldabs*}}

\let\oldnorm\norm
\def\norm{\@ifstar{\oldnorm}{\oldnorm*}}
\makeatother

\begin{document}

\title{Backreaction and order reduction in initially contracting models of the universe }

\author{Leda Gao}
\email{gaol18@wfu.edu}
\affiliation{Department of Physics, Wake Forest University, Winston-Salem, North Carolina 27109, USA}
\author{Paul R. Anderson}
\email{anderson@wfu.edu}
\affiliation{Department of Physics, Wake Forest University, Winston-Salem, North Carolina 27109, USA}
\author{Robert S. Link}
\affiliation{Department of Physics, Wake Forest University, Winston-Salem, North Carolina 27109, USA}

\begin{abstract}
The semiclassical backreaction equations are solved
in closed Robertson-Walker spacetimes containing a positive cosmological constant and a conformally coupled massive scalar field.
Renormalization of the stress-energy tensor results in higher derivative terms that can lead to solutions that
vary on much shorter time scales than the solutions that would occur if the higher derivative terms were not present.  These extra solutions can be eliminated through the use of order reduction.  Four different methods of order reduction are investigated.
These are first applied to the case when only conformally invariant fields, with and without classical radiation, are present.  Then they are applied to the massive conformally coupled scalar field.  The effects of different adiabatic vacuum states for the massive field are considered.
It is found that if enough particles are produced, then the Universe collapses to a final singularity.  Otherwise it undergoes a bounce, but at a smaller value of the scale factor (for the models considered) than occurs for the classical de Sitter solution.
The stress-energy tensor incorporates both particle production and vacuum polarization effects.  An analysis of the energy density of the massive field is done to determine when the contribution from the particles dominates.
\end{abstract}

\maketitle

\section{Introduction}
\label{sec-intro}

Quantum field theory in curved space predicts interesting phenomena such as vacuum polarization and particle production.  Semiclassical gravity is an important way to account for the effects of quantized fields on the spacetime geometry.  The place where semiclassical gravity has been explored the most thoroughly is cosmology,
where the homogeneity and isotropy of the very early universe, or at least the part of it that became the present observable universe, makes the equations much more tractable.

Most of the explorations of quantum effects in the early universe have been in the context of models in which the universe is expanding, including inflationary models.  Once the universe begins to decelerate, quantum effects become smaller in time.  However, there are interesting models in which the Universe initially contracts and then undergoes a bounce at a small but nonzero value of the scale factor.  (For reviews see e.g.~\cite{bounce-review,brandenberger-peter,battefeld-peter,novello-bergliaffa}.)
One feature of most contracting models is that quantum effects are expected to become more important as the bounce is approached.

For many models the bounce occurs at a scale comparable to the Planck scale where semiclassical gravity is not a valid approximation.  However, there are simple models in which a bounce naturally occurs at length scales well above the Planck scale and energy scales well below it. One of these is de Sitter space.  In closed cosmological coordinates (which cover the entire manifold) de Sitter space begins with an infinite size, contracts down to a minimum size that depends on the inverse value of the positive cosmological constant, and then expands to an infinite size.  It thus provides a laboratory in which the effects of semiclassical gravity can be studied.  Further, as has been pointed out in~\cite{dS-bounce}, one can have a model in which there is an effective cosmological constant that is very large so that the Universe contracts to a scale comparable to the GUTs scale and then enters an inflationary phase as it begins expanding.

In this paper we study quantum effects in de Sitter space by solving the semiclassical backreaction equations for a massive conformally coupled scalar field when the cosmological constant is positive and the field is in an adiabatic vacuum state~\cite{b-d-book}.  The Bunch-Davies state for this field results in a stress-energy tensor that is de Sitter invariant~\cite{dowker-critchley} and thus the semiclassical backreaction equations have de Sitter space as an exact solution.  However, this is not the case for any other adiabatic vacuum state.

Studies of quantum effects for massive conformally coupled scalar fields in de Sitter space in closed Robertson-Walker coordinates were previously done in the context of a background field approximation where backreaction effects are ignored~\cite{am-particle,am-vacuum}.  It was shown that for adiabatic vacuum states particle production occurs and as the scale factor approaches its minimum value, the stress-energy of the particles can become significant.  The conclusion was that de Sitter space can be unstable for such states.  Here we solve the semiclassical backreaction equations and validate that prediction.

Higher derivative terms occur in the stress-energy tensor of a quantum field in most four-dimensional spacetimes due to the renormalization process.  In the context of a background field expansion there is no problem with such terms.  However, if the semiclassical backreaction equations are solved,
these higher derivative terms can lead to a variety of extra solutions which in cosmology, depending on the values of the renormalization parameters, can lead to solutions that expand or contract extremely rapidly~\cite{fhh,and-1,and-2,azuma-wada} and are generally thought to be unphysical.
Various methods have been used to eliminate these spurious solutions, see e.g.~\cite{flanagan-wald} and references therein.

One of the methods previously used when only conformally invariant fields are considered is that of order reduction.  This was originally
applied to cosmology in~\cite{parker-simon} where the classical Einstein equations were solved and then used to eliminate the higher derivative terms.  Perturbation theory was also used to keep the calculations within the context of the one loop approximation.
Order reduction without the addition of perturbation theory was used in~\cite{roura-verdaguer}, where it was pointed out that perturbation theory
can miss certain secular effects.  Going beyond the one loop approximation can be justified by working within the context of a large $N$ expansion~\cite{Large-N} where $N$ is the number of identical quantum fields.

To our knowledge, order reduction has not been used in cosmology when conformally noninvariant fields are present and particle production occurs.  Particle production is a nonlocal phenomenon, and in most cases ultraviolet divergences occur in the nonlocal terms which makes renormalization nontrivial.  As shown in~\cite{and-3}, the renormalized stress-energy tensor for a conformally coupled massive scalar field in a Robertson-Walker spacetime has the unique property that its higher derivative terms are all local and exactly the same as those for a massless conformally coupled scalar field.  This makes it possible to obtain a straightforward generalization
of previous order reduction techniques to this case.  Further, particle production effects are confined to the nonlocal contribution to the stress-energy tensor, although this part can also contain vacuum polarization effects since the two cannot usually be completely separated in dynamical spacetimes.

For the massive conformally coupled scalar field, the nonlocal contribution to the stress-energy tensor also has no derivatives of the metric associated with it. The latter property allows one to have an ultraviolet finite stress-energy tensor for zeroth order adiabatic states.  In general, only fourth order adiabatic states result in stress-energy tensors that are ultraviolet finite.
This property makes it straight forward to apply the order reduction technique in~\cite{roura-verdaguer} for zeroth order adiabatic states and this is our first method.  It is not obvious how to apply that technique if higher order adiabatic states are chosen and/or if scalar fields with nonconformal coupling to the scalar curvature are chosen.

We also consider three other methods of order reduction which all involve iteration.  The second one consists of first solving the semiclassical backreaction equations using the nonlocal contribution to the stress-energy tensor (which is separately conserved) but not the higher derivative terms.  Then the higher derivative terms
are computed in the background of this solution and used as source terms for the semiclassical equations. Iterations can be done until the desired level of convergence is reached.

The higher derivative terms are contained in two separately conserved tensors~\cite{and-3}.  One tensor simply contains powers of one derivative of the scale factor and thus
does not lead to runaway solutions.  The third method involves solving the semiclassical backreaction equations exactly using this tensor but initially ignoring the second tensor which has terms with multiple derivatives of the scale factor.  Then this second tensor is computed in the background of the solution to the semiclassical equations and used as a source term for a first iteration.  For the second iteration, the second tensor is computed in the background spacetime obtained from the first iteration and used as a source term.  One can continue this procedure until the desired convergence of the solution is achieved.

The fourth method was suggested by Agullo~\cite{agullo}.  First the classical Einstein equations are solved.  Then the stress-energy tensor for the quantum field is computed in that background and used a source term for a first iteration.  The stress-energy tensor for the quantum field
is next computed in this background and used as a source term for a second iteration.  The
procedure is repeated until the desired level of convergence has been achieved.

In Sec.~\ref{sec:QFT}, quantum field theory for a scalar field in a closed Robertson-Walker universe is reviewed along with the method of adiabatic regularization.  In Sec.~\ref{sec:order-reduction}, the four order reduction methods are described in detail.  In Sec.~\ref{sec:order-reduction-conformal}, the order reduction methods are illustrated using two examples in which a closed Robertson-Walker universe contains  conformally invariant quantum fields in the conformal vacuum state.  In this case there is no particle production and the exact stress-energy tensor for the fields is known~\cite{b-d-book}.
One example is a  spacetime with a a positive cosmological constant and no classical matter or radiation.  Here the calculations can be done analytically.  The second example adds classical radiation to mimic the effects of the produced particles.
In Sec.~\ref{sec:massive-field}, the four methods are used to solve the semiclassical backreaction equations in the case of a conformally coupled massive scalar field in a closed Robertson-Walker spacetime with a positive cosmological constant when no other matter or radiation is present.
This allows us to better isolate the effects of the produced particles.  A detailed analysis of the energy density of the massive conformally coupled scalar field and its relationship to particle production is given in Sec.~\ref{sec:particle-production}. Sec.~\ref{sec:summary} contains a summary of our results. A review of the WKB approximation along with our method to obtain adiabatic vacuum states is given in Appendix~\ref{app:adiabatic-states}.  A discussion of the solutions to the classical Einstein equations when classical radiation and a positive cosmological constant are present is given in Appendix~\ref{app:classical}.
Throughout we use units such that $\hbar = G = c = 1$ and our conventions are those of Ref.~\cite{MTW}.

\section{Quantum field theory in a closed universe}
\label{sec:QFT}

The metric for a closed Robertson-Walker universe can be written as
\begin{equation}
ds^2 = a(\eta)^2(-d \eta^2+\frac{d r^2}{1-r^2}+r^2 d\Omega^2)
\end{equation}

A free scalar field with arbitrary mass and conformal curvature coupling satisfies the equation.
\begin{equation}
    \Box\phi +m^2 \phi +\frac{1}{6} R \phi =0
\end{equation}
 The scalar field can be expanded in terms of mode functions.
 \begin{equation}
     \phi = \frac{1}{a(\eta)}\sum_{k,l,m} [a_k Y_k(\boldsymbol{x}) \psi_k(\eta)+a_k ^\dagger Y_k^\ast(\boldsymbol{x})\psi_k^\ast(\eta)]
 \end{equation}

\bes \bea
     && \Delta^{(3)} Y_k(\boldsymbol{x}) = - (k^2-1) Y_k(\boldsymbol{x}) \;, \\
    && \psi_k''+[k^2+m^2a^2]\psi_k = 0 \;.  \label{mode-eq-1}
\eea \ees
The unrenormalized energy density and trace of the stress-energy tensor are
\bes \bea
    \langle \rho \rangle_u &=& - \langle 0 | {T_\eta}^\eta | 0  \rangle_u  = \frac{1}{4\pi^2a^4}\sum_{k=1}^{\infty}k^2[|\psi_k '|^2+(k^2+m^2a^2)|\psi_k|^2] \;, \\
    \left \langle 0 \middle | T \middle | 0 \right \rangle_u\ &=&  -\frac{1}{2\pi^2a^4}\sum_{k=1}^{\infty}k^2m^2a^2|\psi_k|^2
\eea \label{Tab-eta} \ees

If we make the change of variable
\be \psi_k=\frac{f_k}{\sqrt{a}} \, \label{f-def} \ee
and change from conformal time to proper time using the relation $dt = a d \eta$, then the mode equation~\eqref{mode-eq-1} becomes
\begin{equation}
 \ddot{f_k}+(\frac{1}{4}\frac{\dot{a}^2}{a^2}-\frac{1}{2}\frac{\ddot{a}}{a}+\frac{k^2}{a^2}+m^2)f_k=0
\end{equation}
and the energy density and trace are
\bes \bea
  \langle \rho \rangle_u &=& \langle 0  | T_{tt} | 0 \rangle_u  = \frac{1}{4\pi^2a^4}\sum_{k=1}^{\infty}k^2 \left[ a|\dot{f_k}|^2-\frac{1}{2}\dot{a}(f_k\dot{f_k^{\ast}}+f_k^{\ast}\dot{f_k}) \right. \nonumber \\
&& \left. \;  +\frac{|f_k|^2}{a}(k^2+m^2a^2+\frac{\dot{a}^2}{4})\right]  \label{rho-t} \\
\left \langle 0 \middle | T \middle | 0 \right \rangle_u\ &=& -\frac{1}{2\pi^2a^5}\sum_{k=1}^{\infty}k^2m^2a^2|f_k|^2
\label{T-t} \eea \label{Tab-t} \ees

\subsection{Adiabatic Regularization}
\label{sec:adiabatic-regularizaton}

 We use adiabatic regularization~\cite{parker, parker-fulling, fulling-parker, fulling-parker-hu} to renormalize the stress-energy tensor. It has proved to be a useful method for scalar fields in homogeneous cosmological spacetimes. The renormalized stress-energy tensor is
\begin{equation}
    \langle T_{a b} \rangle_r = \langle T_{a b} \rangle_u-\langle T_{a b} \rangle_{ad} \;.  \label{Tab-r-first}
\end{equation}

As shown in~\cite{anderson-parker}, the adiabatic counterterms $\langle T_{a b} \rangle_{ad}$ consist of an integral rather than a sum. This introduces some difficulties when calculating the renormalized stress-energy tensor.  These are overcome using the method in~\cite{anderson-eaker} which we briefly describe next.

First, a high frequency WKB approximation is used to completely isolate the ultraviolet divergence terms in the stress-energy tensor.
Then the difference between the unrenormalized stress-energy tensor and the high frequency WKB approximation to that tensor is computed.  The approximate stress-energy tensor is then added back and the adiabatic counter terms are subtracted from it:
\bes \bea
    \langle T_{a b} \rangle_r &=& \langle T_{a b } \rangle_n+\langle T_{a b} \rangle_{an} \\
    \langle T_{a b} \rangle_n&=&\langle T_{a b} \rangle_u-\langle T_{a b} \rangle_d \\
    \langle T_{a b} \rangle_{an}&=&\langle T_{a b} \rangle_d-\langle T_{a b} \rangle_{ad}
\eea \label{Tab-renorm-method} \ees
The components $\left \langle   T_0^{\hspace{0.1cm}0}   \right \rangle_d$ and $\langle T \rangle_d$ are given by the expressions
\bes \bea
    \left \langle  T_{tt}  \right \rangle_d &=&  \frac{1}{4\pi^2a^4}\sum_{k=1}^{\infty} k^2 \left[ k+\frac{m^2a^2}{2k}-\frac{m^4a^4}{8k^3} \right] \label{rho-d}  \\
    \langle T \rangle_d &=&  -\frac{1}{4\pi^2a^4}\sum_{k=1}^{\infty} k^2 \left[ \frac{m^2a^2}{k}-\frac{m^4a^4}{2k^3} \right]
\eea \label{Td-eqns} \ees The quantity $ \langle   T_{\mu \nu} \rangle_{n}$ must usually be computed numerically.  The quantity
 $ \langle   T_{\mu \nu} \rangle_{an}$ was computed analytically in
~\cite{anderson-eaker} with the result
\bes \bea
      \left \langle   T_{tt}  \right \rangle_{an} &=&\frac{1}{2880\pi^2}\left(-\frac{1}{6}\,{}^{(1)}H_{tt}+ \,^{(3)} H_{tt}\right)-\frac{m^2}{288\pi^2}G_{tt}  \nonumber \\
       & & -\frac{m^4}{64\pi^2}\left[\frac{1}{2}+\log(\frac{\mu^2 a^2}{4})+2C \right]  \\
      \left \langle  T  \right \rangle_{an} &=&  \frac{1}{2880\pi^2}\left(-\frac{1}{6} \, {}^{(1)} H_c^{\hspace{0.1cm}c}+{}^{(3)} H_c^{\hspace{0.1cm}c}\right)-\frac{m^2}{288\pi^2}G_c^{\hspace{0.1cm}c}   \nonumber \\
        && + \frac{m^4}{16\pi^2}\left[1+\log(\frac{\mu^2 a^2}{4})+2C\right] \;. \eea \label{Tab-analytic} \ees
Here C is Euler's constant, $\mu=m$ for a massive field, $G_{a b}$ is the Einstein tensor,
\bes \bea  ^{(1)\!}H_{ab} &=& - \frac{1}{\sqrt{-g}} \frac{\delta}{\delta g^{ab}} \int d^4 x \, \sqrt{-g} \, R^2 \nonumber \\
& =&  - 2 g_{ab} \Box R + 2 \nabla_a \nabla_b R - 2 R R_{ab}
+ \frac{1}{2} g_{ab} R^2 \;, \label{H1-def} \\
   ^{(3)\!}H_{ab} &=& -R_a^c R_{c b} + \frac{2}{3} R R_{ab} + \frac{1}{2} g_{ab} R_{cd} R^{cd} - \frac{1}{4} g_{ab} R^2 \;. \label{H3-def} \eea \label{H1-H3-def}  \ees
Two independent components of these tensors are
\bes \bea
G_{tt} &=& \frac{3\dot{a}^2}{a^2}+\frac{3 k}{a^2} \label{Gtt} \\
G_a^{\hspace{0.1cm}a}&=& -\frac{6\ddot{a}}{a}-\frac{6\dot{a}^2}{a^2}-\frac{6 k}{a^2} \;. \label{G}  \\
    {}^{(1)} H_{t t}&=& -\frac{36\dddot{a}\dot{a}}{a^2}+\frac{18\ddot{a}^2}{a^2}-\frac{36\ddot{a}\dot{a}^2}{a^3}+\frac{54\dot{a}^4}{a^4}+\frac{36 k \dot{a}^2}{a^4}-\frac{18 k^2 }{a^4}
  \\
    {}^{(1)} H_a^{\hspace{0.1cm}a}&=& \frac{36\ddddot{a}}{a}+\frac{108\dddot{a}\dot{a}}{a^2}+\frac{36\ddot{a}^2}{a^2}-\frac{180\ddot{a}\dot{a}^2}{a^3}-\frac{72 k \ddot{a}}{a^3}
  \\
    {}^{(3)} H_{tt}&=& \frac{3\dot{a}^4}{a^4}+\frac{6 k \dot{a}^2}{a^4}+\frac{3 k^2 }{a^4}
  \\
    {}^{(3)} H_a^{\hspace{0.1cm}a}&=& -\frac{12\ddot{a}\dot{a}^2}{a^3}-\frac{12 k \ddot{a}}{a^3}
\eea \label{1H-3H} \ees
Note that the term in $\la T_{ab} \ra_{an}$ that is proportional to the Einstein tensor could be taken over to
the left hand side of the equations, in which case it would provide a finite renormalization of Newton's constant $G$ which has been set equal
to 1 here.  However, we do not do this because if the scale factor is slowly varying, then one can use a WKB approximation for the exact
modes if the mass of the field is nonzero.  Using a fourth order or higher WKB approximation will cancel all of the terms in $\la T_{\mu \nu} \ra_{an}$.  For this reason we keep this term on the right hand side of the semiclassical backreaction equations.

\section{Order reduction}
\label{sec:order-reduction}

The semiclassical backreaction equations when classical radiation is present are given by
\bes \bea {\left(\frac{\dot{a}}{a}\right)}^2 &=& \frac{8\pi}{3}  \left( \frac{c_r}{a^4}  + \langle \rho \rangle   \right) -\frac{1}{a^2}+\frac{\Lambda}{3} \label{semi-class-rho} \\
       \frac{\ddot{a}}{a} &=& -\frac{1}{a^2}-\frac{\dot{a}^2}{a^2}-\frac{4\pi}{3}\langle T \rangle+\frac{2}{3}\Lambda
    \label{semi-class-T}    \eea \label{semi-class-eqs} \ees
where $c_r a^{-4}$ is the energy density of the classical radiation.

After adiabatic regularization is applied to the stress-energy tensor, Eq.~\eqref{semi-class-rho} contains terms with up to three derivatives of the scale factor and~\eqref{semi-class-T} contains terms with up to four derivatives of the scale factor.
This can result in runaway solutions and solutions that don't grow quickly but which vary on very short timescales, both of which seem unlikely to be physically relevant.\footnote{Because general relativity is expected to be a low energy approximation to some other theory, one expects
that there will be higher derivative terms in the gravitational Lagrangian.  In a conformally flat spacetime, the leading order term in the
semiclassical backreaction equations would be of the form $\alpha \, ^{(1)}H_{\mu \nu}$.  The dimensionless constant $\alpha$ must be fixed by experiment or observation.  Starobinsky inflation~\cite{star-infl-1,star-infl-2} can occur if the value of $\alpha$ is of order $10^9$~\cite{huang}.
Starobinsky inflation requires a solution that grows rapidly during the inflationary phase and during the reheating phase undergoes rapid oscillations in the scalar curvature $R$, although the scale factor itself grows monotonically.  Here we set the value of $\alpha$ to zero and only include
the contribution from $^{(1)}H_{\mu \nu}$ that comes from the stress-energy tensors of the quantum fields. }
      The purpose of order reduction techniques is to find ways to eliminate such solutions and focus on those that are better behaved.  Next we describe in detail the four methods of order reduction that we use in the following sections.

The order reduction techniques developed in~\cite{parker-simon,roura-verdaguer} were applied to the case of conformally invariant fields in Robertson-Walker spacetimes which are conformally flat.  Thus the stress-energy tensor is known analytically.  For a massive conformally coupled scalar field, part of the stress-energy tensor is nonlocal and in most cases must be computed numerically.  There are no derivatives of the scale factor that explicitly occur in this part.  The other part is the same as the
stress-energy tensor for the conformally invariant scalar field in the conformal vacuum state.
The fact that there are  no derivatives of the scale factor in the nonlocal part makes it straight-forward to adapt the approach in~\cite{roura-verdaguer} to this case.  Our first method involves such an adaptation.  We also investigate two other methods developed by us, and one due to Agullo~\cite{agullo}.

To understand the first method, it is useful to begin with a review of the Parker-Simon approach.
 In their method, the order $\hbar$ quantum corrections to the semiclassical backreaction equations are first omitted so that one is
 working only with the classical terms. The classical Einstein equations contain terms up to and including second-order derivatives of the scale factor. For a homogenous and isotropic universe, one component has a term containing first order derivatives and it can thus be solved to obtain the first order derivative of the scale factor in terms of quantities such as the energy density and the cosmological constant.  This equation can be differentiated to obtain an equation for the second derivative of the scale factor.  In this way one can obtain expressions for the third and fourth derivatives.  When these are substituted into the equation for the stress-energy tensor and used repeatedly, the result is an expression that contains at most terms with one derivative of the scale factor.
  The stress-energy tensor for the quantum
 field is of order $\hbar$ so they solve the resulting equations using perturbation theory, keeping only terms up to order $\hbar$.

It was pointed out in~\cite{roura-verdaguer} that perturbation theory can miss secular effects.  An alternative method was proposed in which the order reduction is done in the same way as in the Parker Simon approach but the equations are solved exactly rather than with the use of perturbation theory.  This procedure can be justified if one works in the context of a large $N$ expansion with $N$ the number of identical quantized fields.  The reason is that the semiclassical backreaction equations appear at leading order in such an expansion.  In~\cite{roura-verdaguer} this method was used to solve the semiclassical backreaction equations for perturbations of de Sitter space when conformally invariant fields are present.

Our first method of order reduction, which we denote by M1, is very similar to this.
If the quantities $\langle \rho \rangle$ and $\langle T \rangle$ in the semiclassical back-reaction equations~\eqref{semi-class-eqs} are omitted, then the equations for the first four derivatives of the scale factor are:
\bes \bea
    \dot{a}^2 &=& \frac{8 \pi c_r}{3 a^2} -1+\frac{\Lambda a^2}{3} \\
    \ddot{a} &=& - \frac{8 \pi c_r}{3 a^3} +  \frac{\Lambda a}{3} \\
    \dddot{a} &=&  \frac{8 \pi c_r \dot{a}}{ a^4} +  \frac{\Lambda\dot{a}}{3} \\
    \ddddot{a} &=&  - \frac{320 \pi^2 c_r^2}{3 a^7} + \frac{32 \pi c_r}{a^5} - \frac{80 \pi c_r \Lambda}{9 a^3} +                     \frac{\Lambda^2a}{9}
\eea \label{order-reduc-eqns} \ees
If these expressions are substituted into Eqs.~\eqref{1H-3H} and the result is substituted into~\eqref{Tab-analytic}, then  the result is that
$\langle \rho \rangle_r$ and $\langle T \rangle_r$ become functions of the scale factor but not its derivatives.
Substituting the order-reduced stress-energy tensor into
 the semiclassical backreaction equations~\eqref{semi-class-eqs} then results in equations that are of the same order in terms of derivatives of the scale factor as the classical Einstein equations.

 Unlike the next three methods which all involve iteration, this method does not so long as
 a zeroth order adiabatic state is used for the massive scalar field.  If a second or higher order adiabatic state is used, then
 at the initial time step there are derivatives of the scale factor on the right hand side of the equations and this method cannot be used, at least in its simplest form.
 Note also that the term in $\la T_{ab} \ra_{an}$ that is proportional to $G_{ab}$ remains on the right hand side of the equations and
 thus the order reduction process is used on it as well.  This results in finite renormalizations of the radiation, spatial curvature, and
 cosmological constant terms on the right hand side of the semiclassical backreaction equations.

For the second method, denoted by M2, we first assume that ${}^{(1)} H_\mu^{\hspace{0.1cm}\mu} $ and ${}^{(3)} H_\mu^{\hspace{0.1cm}\mu}$ are much smaller than the other terms in  $ \left \langle  T  \right \rangle_{an}$ for the solutions of interest.  Then, we solve Eqs.~\eqref{semi-class-eqs} numerically for $a(t)$ without including these terms.  We next fit the numerical data for $a$, $\dot{a}$ and $\ddot{a}$ to polynomial functions of the time $t$. The functions for  $\dddot{a}$ and $\ddddot{a}$ are obtained by taking derivatives of the fit for $\ddot{a}$.  Then, we used these functions to evaluate ${}^{(1)} H_\mu^{\hspace{0.1cm}\mu} $ and ${}^{(3)} H_\mu^{\hspace{0.1cm}\mu}$.  Next, Eqs.~\eqref{semi-class-eqs} are solved again but with ${}^{(1)} H_\mu^{\hspace{0.1cm}\mu} $ and ${}^{(3)} H_\mu^{\hspace{0.1cm}\mu}$ included as source terms.  For the next iteration, the numerical data for the scale factor and its first two derivatives is again fitted to polynomial functions of the time, ${}^{(1)} H_\mu^{\hspace{0.1cm}\mu} $ and ${}^{(3)} H_\mu^{\hspace{0.1cm}\mu}$ are computed from these fits and used as source terms in the semiclassical backreaction equations.  The process is repeated until the scale factor $a$ converges to the desired accuracy.

 For the third method, denoted by M3, the ${}^{(1)} H_\mu^{\hspace{0.1cm}\mu} $ term in Eqs.~\eqref{semi-class-eqs} is treated as a source term while the ${}^{(3)} H_\mu^{\hspace{0.1cm}\mu}$ term is treated the same as all of the other terms.  This tensor has no terms containing either
 $\dddot{a}$ or $\ddddot{a}$.  Then~\eqref{semi-class-rho} is solved for $\dot{a}$ and~\eqref{semi-class-T} is solved for $\ddot{a}$.  The resulting equations are then solved with ${}^{(1)} H_\mu^{\hspace{0.1cm}\mu} $ set equal to zero.
  Then the same procedure is used to evaluate ${}^{(1)} H_\mu^{\hspace{0.1cm}\mu} $ as is used in M2 and the result is used as a source term for the first iteration. The rest of the iterations proceed in the same fashion as M2 except that only ${}^{(1)} H_\mu^{\hspace{0.1cm}\mu} $  is treated as a source term.

The fourth method is due to Agullo~\cite{agullo}.  For this method, denoted by M4, first the classical Einstein equations are solved.  Then  the full expressions for  $\langle \rho \rangle$ and $\langle T \rangle$ are evaluated in the resulting geometry.  For the nonlocal parts, this involves first solving the mode equation in that background.  For the first iteration, Eqs.~\eqref{semi-class-eqs} are solved with these terms used as source terms.  Then, $\langle \rho \rangle$ and $\langle T \rangle$ are computed in the new background geometry and used as a source for the next iteration.  This process continues until the iterations converge to the desired accuracy.

All four methods described above work for zeroth-order adiabatic states. M2, M3, and M4 work for second order adiabatic states.  However, only M4 works for a fourth-order or higher adiabatic vacuum state.  Note that M4 is very robust and should work for nonconformally coupled scalar fields as well.  For such fields, $\la T_{\mu \nu} \ra_d$ contains terms with up to four derivatives of the scale factor~\cite{anderson-eaker}.  Since these are
used to explicitly cancel the divergences in $\la T_{\mu \nu} \ra_u$, methods M1, M2, and M3 will not work, at least not without some significant modification.

\section{Order reduction methods applied to quantized conformally invariant fields.}
\label{sec:order-reduction-conformal}

In this section we apply the order reduction methods described in the previous section to the case of quantized conformally invariant scalar fields in closed Robertson-Walker spacetimes.  Since these are conformally flat spacetimes, the stress-energy tensor for any homogeneous and isotropic state has the form~\cite{wald}
\be \la T_{ab} \ra = T^{cr}_{ab} + \la 0| T_{ab}|0 \ra \;, \ee
where the first term on the right is the stress-energy tensor for classical radiation and the second is the stress-energy tensor if the field
is in the conformal vacuum state.
\be \la 0| T_{ab} | 0 \ra =
 -\frac{1}{6} \alpha_q \; ^{(1)}H_{ab} + \beta_q \; ^{(3)}H_{ab} \;. \label{Tab-conformal-vacuum} \ee
Here
\bes \bea \alpha_q &=& \frac{1}{2880 \pi^2} (N_0 + 6 N_{\frac{1}{2}} + 12 N_1)\,, \label{alphaq} \\
     \beta_q &=& \frac{1}{2880 \pi^2} (N_0 + 11 N_{\frac{1}{2}} + 62 N_1 ) \;, \label{betaq} \eea \label{alphaq-betaq} \ees
with $N_0$ the number of conformally invariant scalar fields, $N_{\frac{1}{2}}$ the number of massless spin $\frac{1}{2}$ fields, and $N_1$ the number of massless spin $1$ fields.

In what follows we consider the case where the fields are in the conformal vacuum state and the case where they are in some other homogeneous and isotropic state.  The latter is equivalent to having the quantum fields in the conformal vacuum state and having classical radiation~\cite{wald}.

\subsection{No classical radiation}
\label{sec:no-class-rad}

The scale factor for de Sitter space in closed cosmological coordinates is
\be a(t) = \frac{1}{H} \cosh H t;, \label{ds-scalefactor} \ee
with $H$ a positive constant.
For this form of the scale factor, one finds that
\bes \bea ^{(1)}H_{ab} &=& 0 \\
          ^{(3)}{H_{tt}} &=& 3 H^4 \\
          ^{(3)}{H_c}^c &=& - 12 H^4   \;.  \eea \label{H-1-3-dS} \ees
If only conformally invariant fields in the conformal vacuum state are present, then de Sitter space is an exact solution to the semiclassical backreaction equations even if there is no cosmological constant~\cite{starobinsky-1980}.
If a cosmological constant is also present, then an exact solution to the semiclassical backreaction equations~\eqref{semi-class-eqs} with $c_r = 0$ is de Sitter space with the scale factor given by~\eqref{ds-scalefactor} and
\bes \bea
   H^2 &=& \left(\frac{\dot{a}}{a}\right)^2 + \frac{1}{a^2}  =  \frac{1}{16 \pi \beta_q} \left( 1 \pm \sqrt{1-32 \pi \beta_q H_0^2}\right) \;, \label{Hexact} \\
   H_0^2 &=& \frac{\Lambda}{3}  \;. \eea \label{HH0} \ees
The solution with the plus sign diverges in the limit $\beta_q \to 0$, which is the limit that quantum effects vanish. So this solution is physically unacceptable.  The solution with the minus sign reduces to the de Sitter solution of the
classical Einstein equations, $H = H_0$ in the limit $\beta_q \to 0$.
For comparison with the solutions from the order reduction methods, it is useful to expand in powers of $\beta_q$ with the result
\be H^2 = H_0^2 + 8 \pi \beta_q H_0^4 +128 \pi^2 \beta_q^2 H_0^6 + \ldots\;. \label{H2-expansion} \ee

As discussed in Sec.~\ref{sec:order-reduction}, the method M1 involves using the classical Einstein equations to replace the higher
derivative terms in the stress-energy tensor for the quantum field.  This results in a stress-energy tensor that is simply a function
of the scale factor. The higher derivative terms for the scale factor when no classical radiation is present are given in~\eqref{order-reduc-eqns} with
$c_r = 0$.  They are the same values one would obtain for de Sitter space when $H = H_0$.
Thus for method (1), the semiclassical backreaction equations are
\be \left(\frac{\dot{a}}{a}\right)^2 = \frac{-1}{a^2} +  H_0^2 + 8 \pi \beta_q H_0^4 \ee
which has the de Sitter solution~\eqref{ds-scalefactor} with
\be H^2 = H_0^2 + 8 \pi \beta_q H_0^4 \;. \label{H-H0-sol} \ee
This is in agreement with the exact solution~\eqref{H2-expansion} to first order in $\beta$.  This result was previously found in~\cite{roura-verdaguer}.

For M2, which in this case is equivalent to M4, the stress-energy tensor for the conformally invariant fields is ignored on the zeroth iteration, which means that the classical Einstein equations are solved.  On each successive iteration we compute the stress-energy tensor for the conformally invariant fields using the metric obtained from the previous iteration and then use this stress-energy tensor as a source term for the semiclassical backreaction equations.  When there is no classical radiation, we get a new de Sitter solution upon each iteration if we only keep terms in the result up to $O(\beta^i)$.  This is reasonable because upon subsequent iterations the coefficients of these terms do not change whereas the coefficients of any higher powers of $\beta$ do change. Letting $H_{i-1}$ be the solution for the previous iteration we have
\be H_i^2 = H_0^2 + 8 \pi \beta_q H_{i-1}^4   \;. \label{H-iter} \ee
Starting with the de Sitter solution to the classical Einstein equations $H = H_0$, the first iteration then gives an expression for $H_1^2$
that is equal to the first two terms on the right in~\eqref{H2-expansion}.  After that the i'th iteration reproduces the expansion~\eqref{H2-expansion} to $O(\beta^i)$ along with higher order terms in $\beta$ that have different coefficients.

Because $^{(1)}H_{ab} = 0$ in de Sitter space, M3 is equivalent to solving the exact semiclassical backreaction equations and yields the solution~\eqref{Hexact}.

\subsection{Classical radiation}
\label{sec:class-rad}

When both classical radiation and a positive cosmological constant are present, the solution to the classical Einstein equations depends on the amount of radiation and the size of the cosmological constant.  In a closed universe, the solutions of interest start out with infinite size and either contract to a final singularity or undergo a bounce and expand to an infinite size

The energy density for classical radiation is
\be \rho_{\rm rad} = \frac{c_r}{a^4}  \;, \label{rho-rad} \ee
with $c_r$ a positive constant.  The trace of the stress-energy tensor is zero.

To minimize the number of parameters in the classical solution, it is useful to work with the following scaled variables.\footnote{Note that these variables only simplify the classical solution when classical radiation is present, so we only use them in this section. }
\bes \bea  \alpha &=& \sqrt{\Lambda} \; a  \\
           \tau &=& \sqrt{\Lambda} \; t \; . \eea \ees
The semiclassical backreaction equations when radiation plus conformally invariant fields are present in a closed Roberson-Walker universe are
\bes \bea \left(\frac{\alpha^{'}}{\alpha} \right)^2 &=& \frac{A}{\alpha^4} - \frac{1}{\alpha^2} + \frac{1}{3} + \frac{8 \pi}{3\Lambda} \la 0 |\rho |0 \ra \;, \label{rho-eq} \\
         \frac{\alpha''}{\alpha} &=& - \left(\frac{\alpha'}{\alpha}\right)^2 - \frac{1}{\alpha^2} + \frac{2}{3} - \frac{4 \pi}{3 \Lambda} \la 0| T | 0 \ra \;, \label{T-eq} \\
         \la 0| \rho |0 \ra &=& \Lambda^2 \left\{-\frac{\alpha_q}{6} \left[-\frac{36 \, \alpha''' \alpha'}{\alpha^2}+\frac{18\, (\alpha'')^2}{\alpha^2}-\frac{36 \, \alpha'' (\alpha')^2}{\alpha^3}+\frac{54 \, (\alpha')^4}{\alpha^4}+\frac{36 \, (\alpha')^2}{\alpha^4}-\frac{18 }{\alpha^4} \right] \right.  \nonumber \\
   & & \left.  \;\;\;          + \beta_q \left[\frac{3 \,(\alpha')^4}{\alpha^4}+\frac{6 \, (\alpha')^2}{a^4}+\frac{3 }{\alpha^4} \right]\right\}  \;, \label{rho} \\
          \la 0 |T |0 \ra &=&\Lambda^2 \left\{ -\frac{\alpha_q}{6} \left[\frac{36 \, \alpha''''}{\alpha}+\frac{108 \, \alpha''' \alpha'}{\alpha^2}+\frac{36 \, (\alpha'')^2}{\alpha^2}-\frac{180 \, \alpha'' (\alpha')^2}{\alpha^3}-\frac{72 \, \alpha''}{\alpha^3} \right] \right. \nonumber \\
          & & \left. \;\;\;  + \beta_q \left[ -\frac{12 \, \alpha'' (\alpha')^2}{\alpha^3}-\frac{12\, \alpha''}{\alpha^3} \right] \right\} \label{T} \\
            A &\equiv&  \frac{8 \pi}{3} c_r \Lambda  \;. \label{Adef}
         \eea \label{eqs-class-rad-ci-fields} \ees

The equations for method M1 can be obtained by first transforming the order reduction equations~\eqref{order-reduc-eqns} to scaled variables with the result
\bes \bea
    \alpha^{' \,2} &=& \frac{A}{\alpha^2} -1+\frac{ \alpha^2}{3} \\
    \alpha'' &=& - \frac{A}{\alpha^3} +  \frac{\alpha}{3} \\
    \alpha''' &=&  \frac{3 A \alpha'}{ \alpha^4} +  \frac{\alpha'}{3} \\
    \alpha'''' &=&  - \frac{15 A^2}{\alpha^7} + \frac{12 A }{\alpha^5} - \frac{10 A }{3 \alpha^3} +  \frac{\alpha}{9} \;.
\eea \label{order-reduc-eqns-scaled} \ees
Substituting these into~\eqref{rho} gives
\bes \bea \left(\frac{\alpha^{'}}{\alpha} \right)^2 &=&  \frac{A}{\alpha^4} \left(1 + \frac{16 \pi}{3} \Lambda (2 \alpha_q  + \beta_q) \right) - \frac{1}{\alpha^2} + \frac{1}{3} \left(1 + \frac{8 \pi}{3} \Lambda  \beta_q \right) \nonumber \\
& &  + \frac{8 \pi A^2 \Lambda \beta_q}{\alpha^8} \;. \label{rho-eq-scaled} \eea
Substituting this expression into the first term on the right in~\eqref{T-eq} and then substituting~\eqref{order-reduc-eqns-scaled} into~\eqref{T} gives
\bea  \frac{\alpha''}{\alpha} &=&   - \frac{A}{\alpha^4} \left(1 + \frac{16 \pi}{3} \Lambda (2 \alpha_q  + \beta_q) \right) + \frac{1}{3} \left(1 + \frac{8 \pi}{3} \Lambda  \beta_q \right) \nonumber \\
         & &  - \frac{24 \pi A^2 \Lambda \beta_q}{\alpha^8} \;. \label{Tr-eq-scaled}
\eea \label{eqs-class-rad-ci-fields-scaled} \ees

When only conformally invariant fields and classical radiation are present, then, as mentioned above, M2 and M4 are equivalent.  For M2 we begin by writing
$\alpha = \alpha_0$ and solving~\eqref{rho-eq} and~\eqref{T-eq} with $\la 0|\rho |0\ra = \la 0|T|0 \ra = 0$ to obtain a classical solution.  Then this is substituted into~\eqref{rho} and~\eqref{T}.  For the other terms in~\eqref{rho-eq} and~\eqref{T-eq}, we write $\alpha = \alpha_1$ and solve them for $\alpha_1$.  For a second iteration, $\alpha_1$ is substituted into~\eqref{rho} and~\eqref{T}.  For the other terms in ~\eqref{rho-eq} and~\eqref{T-eq} we write $\alpha = \alpha_2$ and solve them for $\alpha_2$.  In this way one can iterate as many times as desired.  Note that for the first iteration, one can use analytic derivatives of~\eqref{T-eq} with $\la 0|T|0 \ra = 0$ to obtain expressions for
$\alpha_0'''$ and $\alpha_0''''$.  However, for the second iteration, one has only numerical data for $\alpha_1$ and its first two derivatives.  One way to obtain $\alpha_1'''$ and $\alpha_1''''$ is to fit this data to a power series in $\tau$ over the range of times of interest and then compute the derivatives analytically.  This is what is done for the example below.

M3 involves separating out the $\alpha_q$ term from the $\beta_q$ term in~\eqref{rho} and~\eqref{T} and first setting $\alpha_q = 0$.  The resulting second order equations are then solved after setting $\alpha = \alpha_0$.  The result is substituted into the parts of~\eqref{rho} and~\eqref{T} that are proportional to $\alpha_q$ and those parts are used as source terms for the next iteration.  For that iteration, $\alpha = \alpha_1$ is used for the rest of the terms in the backreaction equations.  The iterations proceed in the same way as they do for M2.

\subsubsection{Solutions to the semiclassical backreaction equations}

As shown in Appendix~\ref{app:classical}, the solutions to the classical Einstein equations when classical radiation is present depend upon the value of
$A$ in~\eqref{Adef} and on the initial conditions.  The solutions we are interested in begin with $\alpha = \infty$ at $\tau = -\infty$.
As shown in Appendix~\ref{app:classical}, for $0<A<\frac{3}{4}$ the solutions contract to a minimum nonzero size and then expand again.  For $A = \frac{3}{4}$ the solutions contract to a minimum value of $\alpha = \frac{2}{3}$ in the limit $\tau \to \infty$.  This minimum value of the scale factor is also its value for the static solution in this case, which is the Einstein universe.  For $A > \frac{3}{4}$ the solutions contract to a final singularity.  Thus for a fixed value of $\Lambda$, the late time behavior of the solutions to the classical Einstein equations that begin with $\alpha = \infty$ depend upon the amount of classical radiation that is present.  If there is too much radiation, then they collapse to a final singularity.

When conformally invariant fields are also present, then quantum effects change the detailed behavior of the scale factor as a function of time and they change the specific conditions under which a bounce or collapse to a singularity occurs.  However, we have not found any new qualitative  behaviors when order reduction methods are used.

To illustrate the various order reduction methods in this case, we will focus on bounce solutions to the equations.  Near a bounce we can write a power series solution for the scale factor in terms of $\tau$ if we fix the time so that the bounce is at $\tau = 0$.  Then, since by definition $\alpha' = 0$ at a bounce, we have
\be \alpha = a_0 + a_2 \tau^2 + a_3 \tau^3 + a_4 \tau^4 + \ldots \;. \label{alpha-series} \ee
Substituting this into the exact semiclassical backreaction equations~\eqref{rho-eq} and~\eqref{T-eq}, one finds that $a_0$, $a_2$, and $a_3$ are arbitrary and that for a bounce which by definition is a minimum
\be a_2 = \frac{1}{2} \left[ \frac{1}{8 \pi \Lambda \alpha_q a_0^2} \left(A + 8 \pi \Lambda (\alpha_q + \beta_q) - a_0^2 + \frac{a_0^4}{3}\right) \right]^{1/2} \;. \ee
There can only be bounce solutions if the quantity in square brackets is nonnegative.  This is a rather different condition than the one for bounce solutions in the classical case.  Also in contrast to the classical case, bounce solutions that are not time symmetric exist.

For M1, one can substitute~\eqref{alpha-series} into~\eqref{eqs-class-rad-ci-fields-scaled}.  Then~\eqref{rho-eq-scaled} can be solved for $a_0$ and~\eqref{Tr-eq-scaled} can be solved for $a_2$.  Thus one finds that, as for the classical solutions, $a_0$ and $a_2$ both have fixed values at a bounce.  Computing a time derivative of~\eqref{Tr-eq-scaled}, one sees that $a_3 = 0$.  Computing successive derivatives shows that all bounce solutions in this case are time symmetric.

For M2, one substitutes solutions to the classical equations into~\eqref{rho} and~\eqref{T} and uses them as source terms for the first iteration. In these cases the time at which a bounce solution occurs will usually change on each iteration because the source terms will not vanish at the time of the bounce in the previous iteration.

For M3, one initially finds an exact solution to~\eqref{eqs-class-rad-ci-fields-scaled} when $\alpha_q = 0$ and then for the first iteration uses the terms proportional to $\alpha_q$ as source terms.  The iterations proceed as for M2.

For a single scalar field, backreaction effects are small at scales above the Planck scale.  However, one expects there to be a large number of quantum fields in the early universe, particularly if any of the grand unified theories is correct.  If there are enough quantum fields, then backreaction effects will be more significant.  One can see from~\eqref{eqs-class-rad-ci-fields} that the terms containing $\la \rho \ra$ and $\la T \ra$ in the semiclassical backreaction equations are all multiplied by $\Lambda$.
We consider $N$ identical conformally invariant scalar fields.  Then
\be \Lambda \alpha_q = \Lambda \beta_q = \frac{N \Lambda}{2880 \pi^2} \;. \ee

\begin{figure}[htb!]
\centering
\includegraphics[totalheight=0.3\textheight]{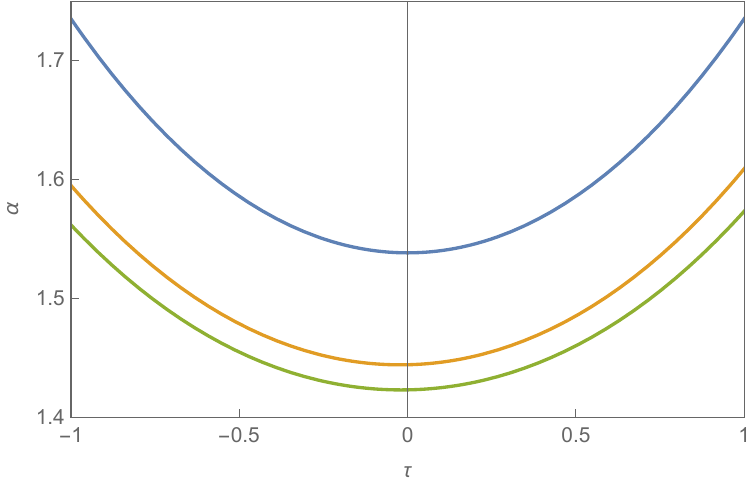}
\caption{Solutions for the scale factor $\alpha$ plotted versus the time $\tau$ using M2. The values of the parameters are $A = \frac{1}{2}$, $N \Lambda = 100$.  From  top to bottom, the curves are for the zeroth (blue), first (orange), and second (green) iterations.  The curves for the first and second iterations have been translated in time so that the minimum is at $\tau \approx 0$.  }
\label{fig:ci-method-2}
\end{figure}

\begin{figure}[htb!]
\centering
\includegraphics[totalheight=0.3\textheight]{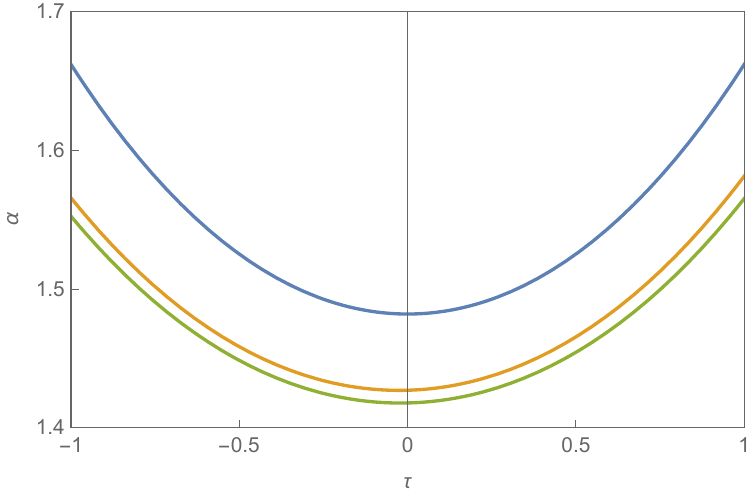}
\caption{Solutions for the scale factor $\alpha$ plotted versus the time $\tau$ using M3.  The values of the parameters are $A = \frac{1}{2}$, $N \Lambda = 100$.   From  top to bottom, the curves are for the zeroth (blue), first (orange), and second (green) iterations.  The curves for the first and second iterations have been translated in time so that the minimum is at $\tau \approx 0$.}
\label{fig:ci-method-3}
\end{figure}

\begin{figure}[htb!]
\centering
\includegraphics[totalheight=0.3\textheight]{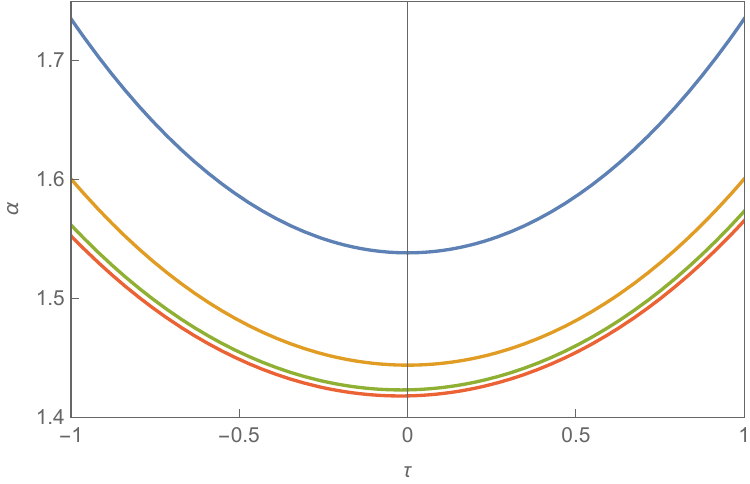}
\caption{Solutions for the scale factor $\alpha$ plotted versus the time $\tau$ using all three methods.  The values of the parameters are $A = \frac{1}{2}$, $N \Lambda = 100$.  From top to bottom, the curves are for the classical solution (blue),  M1 (orange), the second iteration of M2 (green), and the second iteration of M3 (red).  The curves for M2 and M3 have been translated in time so that the minimum of each is at $\tau \approx 0$.}
\label{fig:ci-1}
\end{figure}

For illustration purposes, it is useful to have backreaction effects be somewhat significant.
Thus we consider the case when $N \Lambda = 100$ in Planck units in Figs. \ref{fig:ci-method-2} - \ref{fig:ci-1}.  We also choose $A = \frac{1}{2}$.  In Fig. \ref{fig:ci-method-2}, we show the
solutions for the zeroth, first, and second iterations of M2.  The curve for each iteration has been shifted in time so that the bounce occurs at approximately the same time for all curves.  The iterations appear to be converging.  For the times shown on the plot,
the relative difference between the zeroth and first iterations is of order $10\%$,  while that between the first and second iterations is of order $3\%$ or less.  In Fig.~\ref{fig:ci-method-3} we show the solutions for the zeroth, first, and second iterations for M3.  Again the curves have been shifted so the bounces occur at the same time and again the iterations appear to be converging by about the same amount as for M2.  In both cases, if one extends the plots to larger positive and negative times the relative difference between successive iterations increases.  However, this is at least in part because at large negative times the scale factor is decreasing exponentially and at large positive times it is increasing exponentially.  Since the location of the bounce is only approximately being accounted for, the curves have not been shifted perfectly so one curve starts expanding exponentially earlier than the other and this causes at least some of the increase in the relative difference.

\begin{figure}[htb!]
\centering
\includegraphics[totalheight=0.3\textheight]{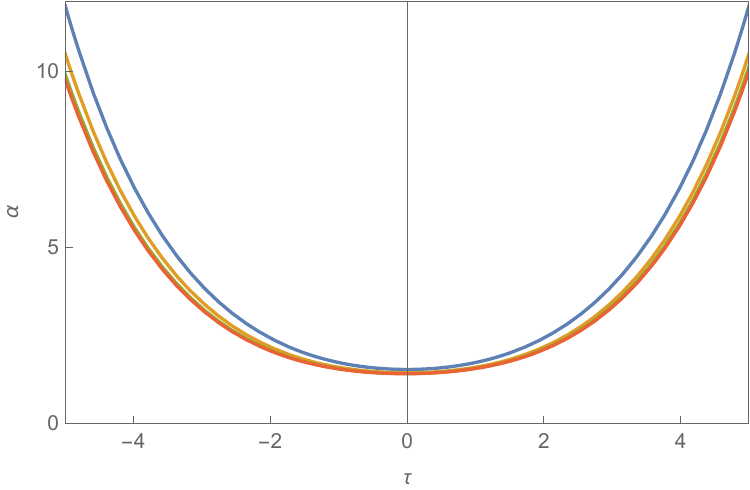}
\caption{Solutions for the scale factor $\alpha$ plotted versus the time $\tau$ for all three methods.  The values of the parameters are $A = \frac{1}{2}$, $N \Lambda = 100$. From  top to bottom curves are for the classical solution (blue),  M1 (orange), the second iteration of M2 (green), and the second iteration of M3 (red).  The curves for M2 and M3 have been translated in time so that the minimum of each is at $\tau \approx 0$.  Note that these curves are so close together that it is difficult to distinguish them on the scale of the plot.}
\label{fig:ci-2}
\end{figure}

The results for the classical solution, the solution for M1, the solution for the second iteration using M2, and the solution for the second iteration using M3 are shown in Figs.~\ref{fig:ci-1} and~\ref{fig:ci-2}, again with the curves for iterated solutions shifted so that the bounce occurs at $\tau \approx 0$.  It can be seen that the bounce occurs at successively smaller values of the scale factor for M1, M2, and M3.  The curves for M2 and M3, which are for the second iteration of each method, are the closest together near the bounce.  However, as can be seen in Fig.~\ref{fig:ci-2}, at larger values of the scale factor, the curves for M1 and M2 are almost indistinguishable on the scale of the plot.

\section{Conformally coupled massive scalar field}
\label{sec:massive-field}

In this section we consider backreaction effects due to a conformally coupled massive scalar field in cases in which the universe is initially contracting and classically will undergo a bounce.  For simplicity, we will not consider any other fields and will not include classical radiation in the problem.  As shown in~\eqref{Tab-renorm-method}, the stress-energy tensor for the field can be broken into
a part that must be computed numerically in most cases along with an analytic part that is shown in~\eqref{Tab-analytic}.  In the limit $m \to 0$ only the first two terms of the analytic part survive if the field is in the conformal vacuum state, which was assumed in the previous section.  Thus, effectively the stress-energy in the massive case has the same terms as for the massless field plus additional terms.

The higher derivative terms in the stress-energy tensor only occur in the terms that survive in the massless limit.  The extra terms induced by the mass do not have higher derivatives associated with them. This results in a distinction between methods M2 and M4 which is not there in the massless case.  In M2, the terms that do not have higher derivatives of the scale factor are kept in the zeroth order iteration because they are separately conserved.  In M4 none of the terms in the stress-energy tensor for the massive scalar field are included for the zeroth iteration.  Then the zeroth order solution is simply that for de Sitter space.  It will be referred to as the dS solution in what follows.

As discussed in the Introduction, we consider adiabatic vacuum states for the conformally coupled massive scalar field.  In general, the vacuum state of the field can be fixed by specifying starting values for the modes $f_k$ and their first derivatives at some starting time $t_1$.  To specify an adiabatic vacuum state, we use the method of adiabatic matching which is discussed in Appendix~\ref{app:adiabatic-states}.  For this method the WKB approximation is used to obtain starting values for the mode functions $f_k$ and $\frac{d f_k}{d t}$ at some initial time $t_1$.\footnote{For the massive scalar field there is no real advantage to using the scaled coordinates that were used in Sec.~\ref{sec:order-reduction-conformal}, so we go back to the original coordinates in which the scale factor $a$ is given as a function of the proper time $t$.} Different truncations of the WKB expansion result in different order adiabatic states.  Different values of $t_1$, for a given truncation of the WKB expansion generate different adiabatic states of the same order.
Because of the conformal coupling, it turns
out that a renormalized stress-energy tensor can be obtained from zeroth and second order adiabatic vacuum states so long as the spacetime is homogeneous and isotropic, which we assume here.  We show results below for zeroth, second, and fourth order adiabatic states.  For scalar fields with nonconformal coupling to the scalar curvature, it is necessary to use a fourth order adiabatic state or higher to obtain a finite stress-energy tensor.

In what follows, we will again take illustrative examples in which the quantum effects are large enough to easily be seen.  The mass for the field is set equal to the Planck mass, thus $m = 1$ in the units we are using.  The cosmological constant is set to $\Lambda = 0.5$ for the zeroth order adiabatic states, to $\Lambda = 1$ for the second order adiabatic states where particle production effects are smaller, and to $\Lambda = 1.5$ for the fourth order adiabatic state for which particle production effects are even smaller.

\subsection{Zeroth order adiabatic states}
\label{zeroth-order-ad}

A zeroth order adiabatic state is the simplest to use for methods M2 and M3, and it is the only adiabatic order for which
we have been able to use M1.  For the calculations we do, the starting values for the modes are obtained using the method outlined in Appendix~\ref{app:adiabatic-states} and Eqs.~\eqref{psi-ad-0} and~\eqref{psiprime-ad-0-1}.
We find solutions for $m=1$ and $\Lambda=0.5$ at the starting times $t_1 = -4$ and $t_1 = -3$.  For the earlier matching time there is enough  particle production for these values of $m$ and $\Lambda$ that the Universe collapses to a singularity instead of bouncing.  For the later matching time, less particle production occurs and the solutions undergo a bounce.  The value of the scale factor at the bounce is smaller than it is for the corresponding classical solution.

\begin{figure}[htb!]
\centering
\includegraphics[totalheight=0.3\textheight]{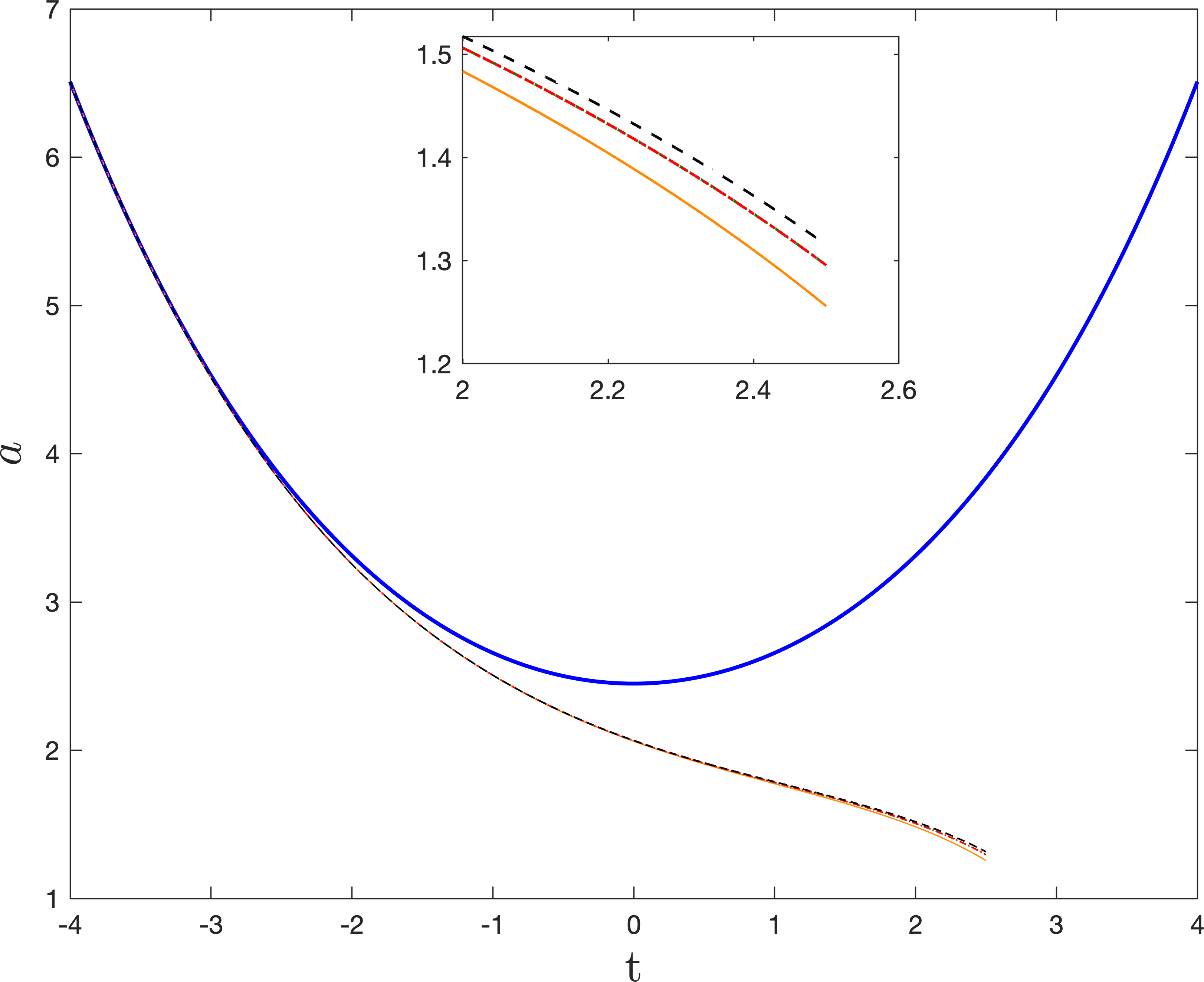}
\caption{Solutions for the scale factor $a$ obtained from all four methods for zeroth order adiabatic vacuum states with matching time $t_1=-4$
are plotted.
The values of the parameters are $m=1$ and $\Lambda=0.5$.  For these values of $m$ and $\Lambda$, and this matching time, there is enough particle production that the scale factor collapses to a singularity.
For the solutions shown, two iterations are implemented for M2, M3, and M4.  There is never an iteration for M1.  The upper curve (blue) is for the classical de Sitter solution. The inset shows the late time behaviors of the rest of the solutions.  From top to bottom, the curves are for the solutions obtained using M4(dashed black), M2(dotted green), M3(dash-dotted red), and M1(solid orange). The curves for the solutions obtained using M2 and M3 overlap and are almost identical. }
\label{fig:t1-4-ad-0-it-2}
\end{figure}

The results for the scale factor for the four methods when $t_1 = -4$ are shown in Fig. \ref{fig:t1-4-ad-0-it-2}. M1 does not include any iterations.  Two iterations are implemented for M2, M3, and M4.  A detailed description of how the iterations are done is given in Sec.~\ref{sec:order-reduction}. The behavior of the scale factor is almost the same for each method before $t\approx 1$. The differences increase as the singularity is approached. Note that the solutions for M2 and M3 are the closest together while those for M1 and M4 have the largest deviation from each other.

\begin{figure}[htb!]
\centering
\includegraphics[totalheight=0.3\textheight]{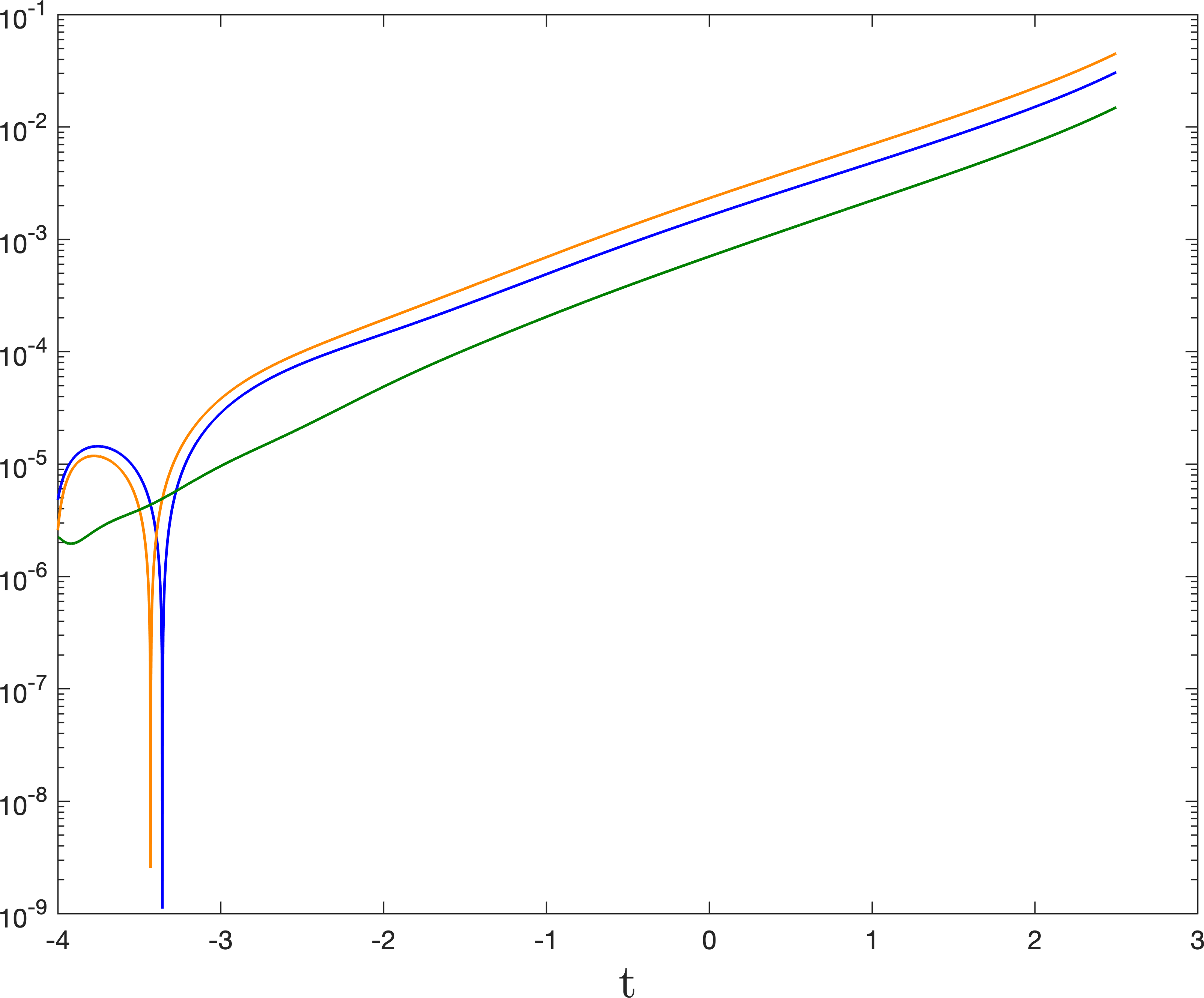}
\caption{The relative differences between the solutions in Fig.~\ref{fig:t1-4-ad-0-it-2} are plotted versus time on a semilog plot.  Differences with the solution obtained using M3 are not included because that solution is almost identical to the one obtained using M2.  From top to bottom, the curves are for the relative differences between the solutions obtained using M1 and M4(orange), M1 and M2(blue), and, M2 and M4(green).  Note that as the singularity is approached, the relative differences increase exponentially. }
\label{fig:t1-4-ad-0-ite-2-rho}
\end{figure}

To better compare the solutions for the different methods,
relative differences between the scale factors are plotted versus time in Fig. \ref{fig:t1-4-ad-0-ite-2-rho}. In the figure, the relative differences are seen to increase exponentially at later times as the singularity is approached.  However, their differences remain small
past the time when the bounce for the classical solution occurs.  The difference between M2 and M3 is omitted because it is significantly smaller than the other differences.

\begin{figure}[htb!]
\centering
\includegraphics[totalheight=0.3\textheight]{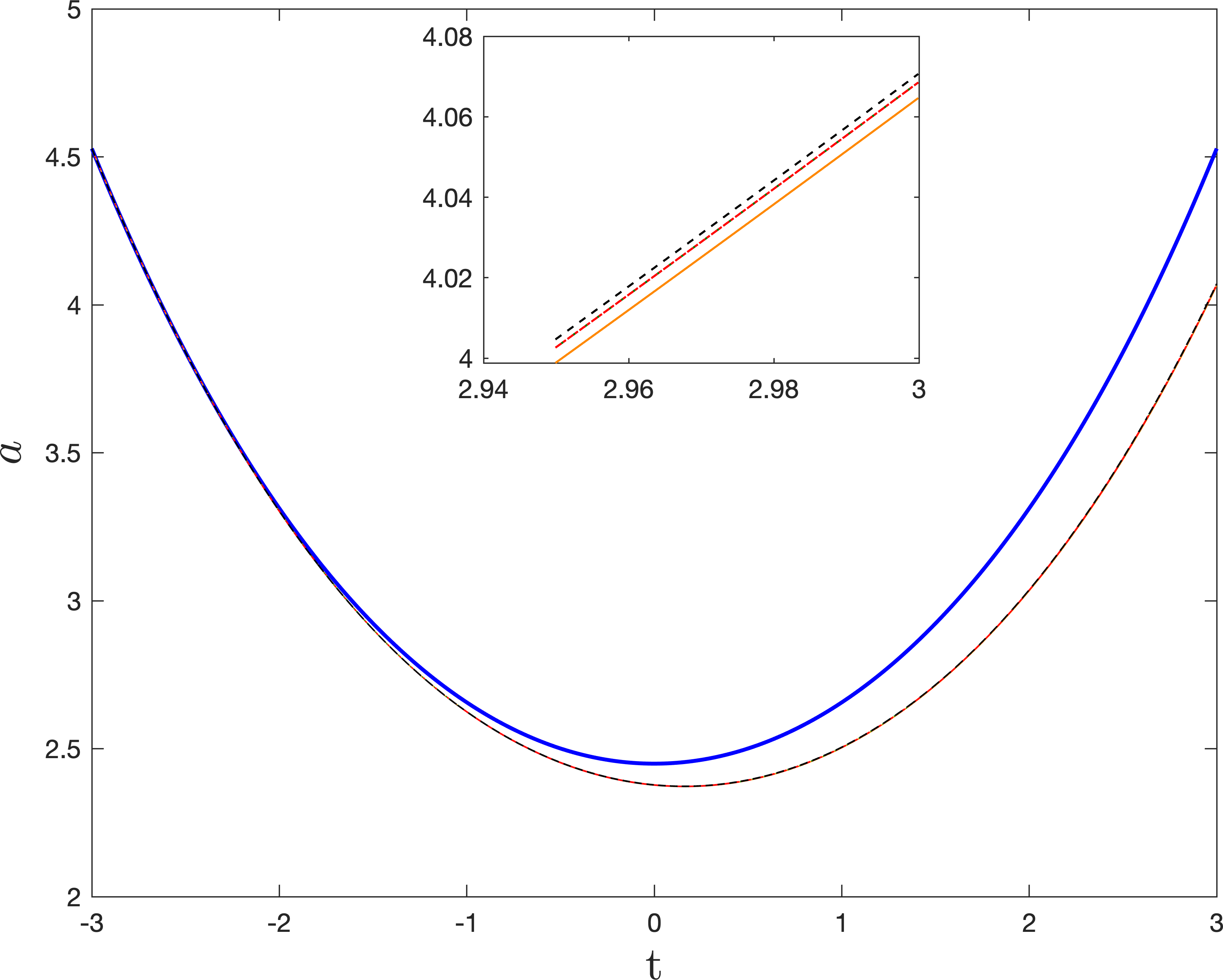}
\caption{Solutions for the scale factor $a$ obtained using all four methods for zeroth order adiabatic vacuum states with matching time $t_1=-3$
are plotted.
The values of the parameters are $m=1$ and $\Lambda=0.5$.  For these values of $m$ and $\Lambda$, and this matching time, the amount of particle production is small enough that the scale factor bounces.
Two iterations are implemented for M2, M3, and M4.
The upper curve (blue) is for the classical de Sitter solution. The inset shows the late time behaviors of the rest of the solutions.
From top to bottom, the curves are for the solutions obtained using M4(dashed black), M2(dotted green), M3(dash-dotted red), and M1(solid orange). The curves for the solutions obtained from M2 and M3 overlap and are almost identical.  The bounces occur for smaller values of the scale factor than for the classical de Sitter solution. }
\label{fig:t1-3-ad-0-it-2}
\end{figure}

The solutions to the semiclassical backreaction equations for the starting time $t_1 = -3$ are shown in Fig. \ref{fig:t1-3-ad-0-it-2}.  Again, there are two iterations for M2, M3, and M4.  As can be seen from the figure, the solutions bounce in this case at smaller values of the scale factor than the corresponding classical de Sitter solution.

\begin{figure}[htb!]
\centering
\includegraphics[totalheight=0.3\textheight]{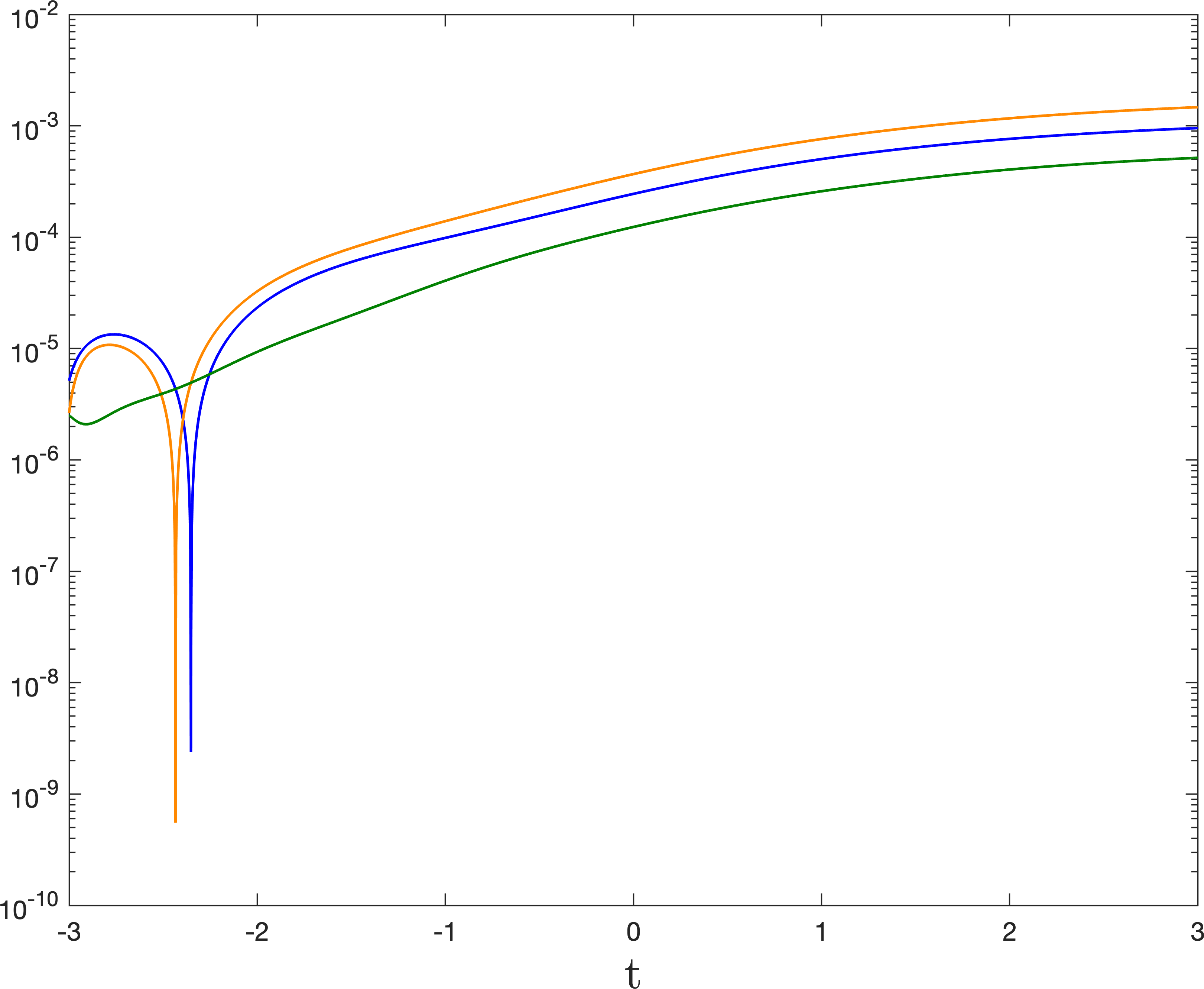}
\caption{The relative differences between the solutions in Fig.~\ref{fig:t1-3-ad-0-it-2} are plotted versus time on a semilog plot.  Differences of solutions with that obtained using M3 are not included because that solution is almost identical to the one obtained using M2.  From top to bottom, the curves are for the relative differences between the solutions obtained using M1 and M4(orange), M1 and M2(blue), and M2 and M4(green). }
\label{fig:t1-3-ad-0-ite-2-rho}
\end{figure}

The relative differences between the scale factors obtained from the different methods are plotted versus time in Fig. \ref{fig:t1-3-ad-0-ite-2-rho}. In general, Fig. \ref{fig:t1-3-ad-0-ite-2-rho} and Fig. \ref{fig:t1-4-ad-0-ite-2-rho} display a similar pattern. However, the relative differences in Fig. \ref{fig:t1-3-ad-0-ite-2-rho} are much smaller than those in Fig. \ref{fig:t1-4-ad-0-ite-2-rho} and do not appear to grow exponentially at late times. One explanation for this is that when the matching time $t_1=-3$ is employed, there is less particle production, causing the scale factor to bounce rather than collapse into a singularity. Quantum effects start to decrease as the Universe expands after the bounce.  So the differences between the solutions begin to level off.

\begin{figure}[htb!]
\centering
\includegraphics[totalheight=0.3\textheight]{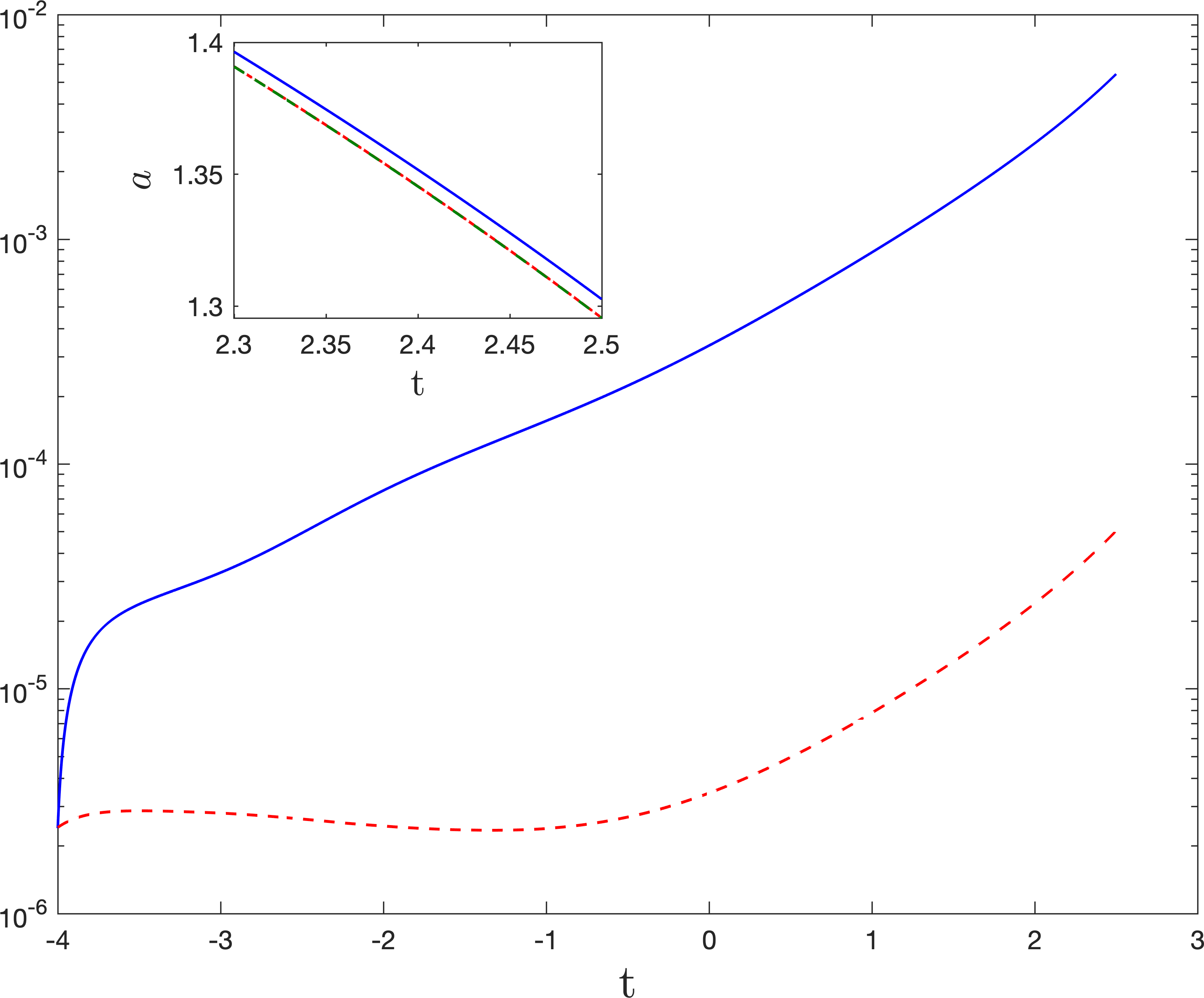}
\caption{The relative differences between different iterations of solutions obtained using M2 are plotted versus time for a zeroth-order adiabatic state in which the adiabatic matching occurs at time $t_1 = -4$. The values of the parameters are $m=1$ and $\Lambda=0.5$. From top to bottom, the curves are for the relative differences between the solutions for the first iteration and no iteration(blue solid), and between the second iteration and the first iteration(red dashed). In the inset, the scale factors for the different iterations at relatively late times are shown.  From top to bottom, the curves are the scale factor for no iterations(solid blue), one iteration(dash-dotted red), and two iterations(dashed green).  The curves for the scale factors for the first and second iterations overlap and are almost identical.}
\label{fig:t1-4-ad-0-compare-iteration}
\end{figure}

In Fig. \ref{fig:t1-4-ad-0-compare-iteration}, the relative differences between the solutions obtained for different iterations of M2 are shown for the case $t_1 = -4$.  It can be seen from the plot that the relative differences increase as the solutions approach the singularity at $\alpha = 0$.  However, the relative difference between the solutions with one and two iterations is significantly smaller than that between the solutions with no iterations and one iteration.  Thus it is clear that the iterations are converging.  Evidence is also found for convergence of the iterations of M3 and M4.

 \subsection{Second-order adiabatic states}

 Solutions to the semiclassical backreaction equations are discussed in this section for
second-order adiabatic states.  For the calculations we do, the starting values for the modes are obtained using the method outlined in Appendix~\ref{app:adiabatic-states} and Eqs.~\eqref{W-first-iteration}.

We do not have a good way to extend M1 to second or higher order adiabatic states.
It is possible to extend M2 and M3 to second order adiabatic states.
What is required is to find self-consistent solutions to~\eqref{semi-class-eqs} at the initial time $t_1$.  The problem is that the terms contained in $\la \rho \ra_n$ when a second order WKB approximation is made for the modes depend on $\dot{a}$.  This is also true
for the parts of $\la \rho \ra_{an}$ that are used for the zeroth iterations for M2 and M3.  Thus it is necessary to find a value for $\dot{a}(t_1)$ that gives a self-consistent solution to~\eqref{semi-class-rho}.  This can be done by starting with the value of $\dot{a}(t_1)$ for the corresponding de Sitter space solution to Einstein's equations and then iterating.  Once the value of $\dot{a}(t_1)$ is determined then a similar procedure can be used to find the self-consistent solution to~\eqref{semi-class-T} at time $t_1$.
For subsequent iterations, this procedure needs to be repeated.  In contrast for M4 there is no problem at any adiabatic order.
One simply uses the geometry from the previous iteration to compute the entire expression for $\la T_{ab} \ra$ and then uses it as a source term for the equations.  Because of this simplicity, we restrict our discussion of second order adiabatic states to M4.

\begin{figure}[htb!]
\centering
\includegraphics[totalheight=0.3\textheight]{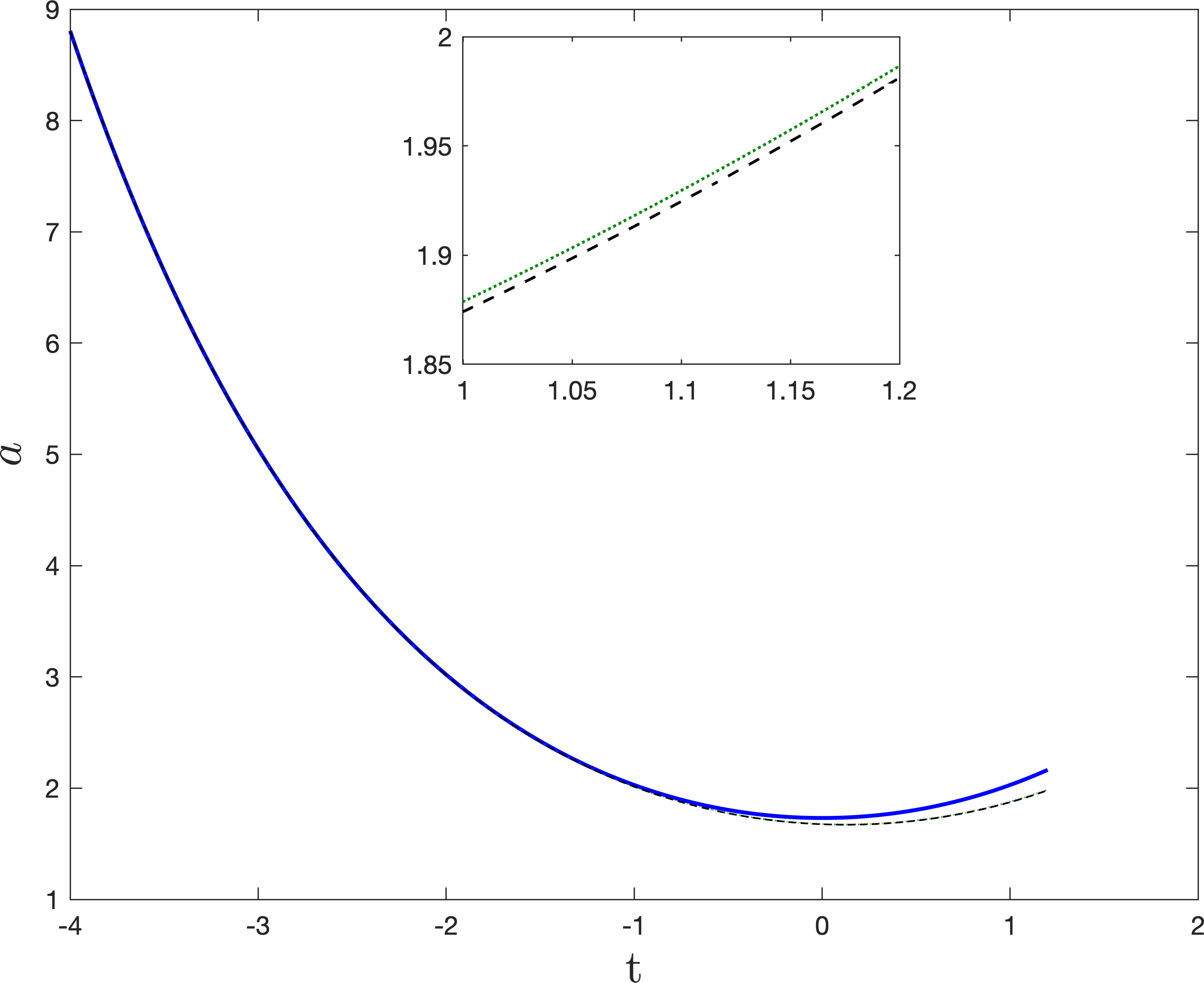}
\caption{Solutions of each iterations for the scale factor $a$ obtained using M4 for a second order adiabatic vacuum state with matching time $t_1=-4$
are plotted along with the classical solution. Two iterations have been carried out for the M4 solution. The values of the parameters are $m=1$ and $\Lambda=1$. The top curve is the classical de Sitter space solution (thick blue) which results from the zeroth iteration.  The middle curve(dotted green) is for the first iteration. The bottom curve (dashed black) is for the second iteration. The curves for the first and second iteration almost overlap. Their difference is shown in the inserted plot. }
\label{fig:t1-4-ad-2-it-2}
\end{figure}

\begin{figure}[htb!]
\centering
\includegraphics[totalheight=0.3\textheight]{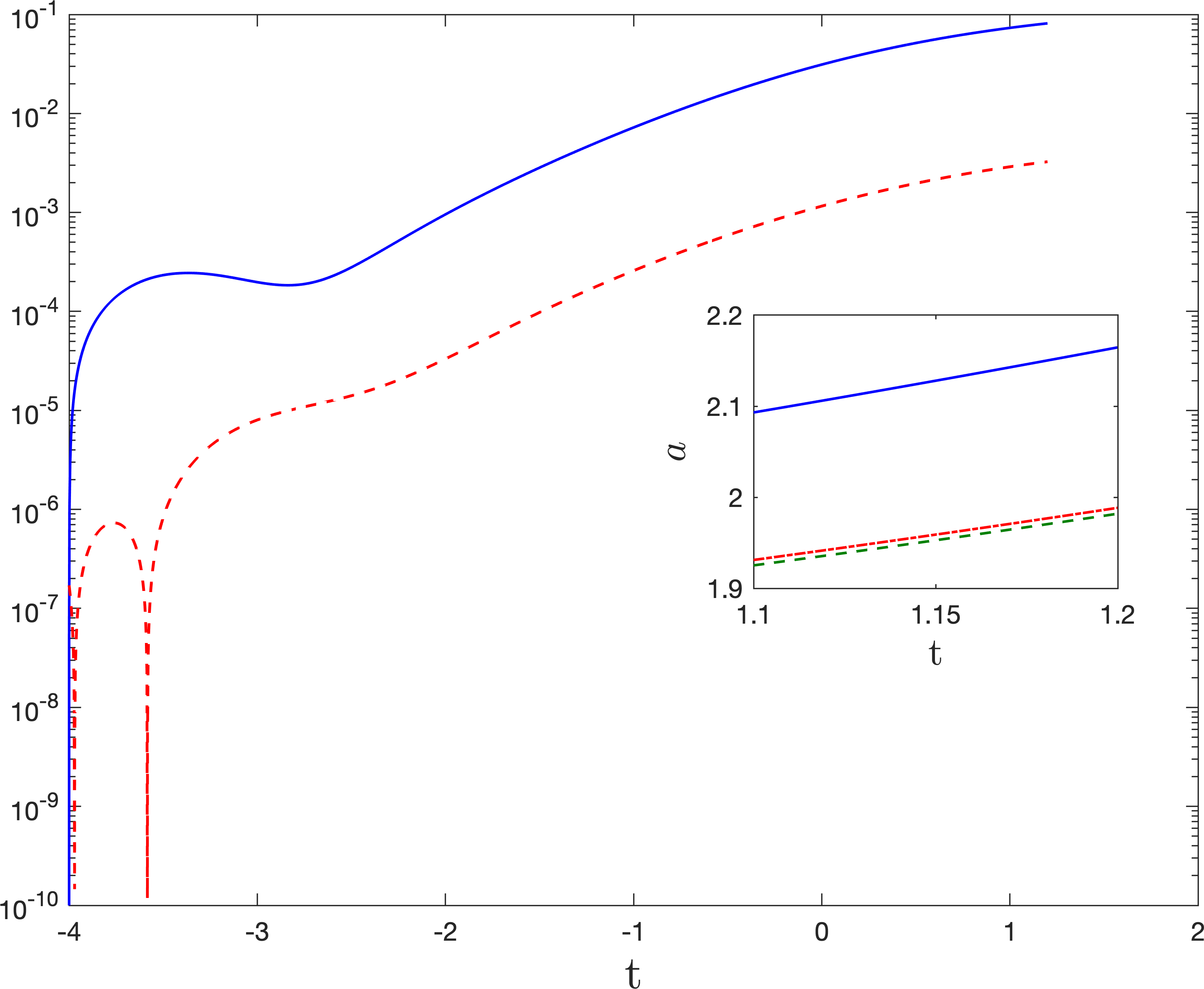}
\caption{The relative differences between different iterations of solutions obtained using M4 for second order adiabatic vacuum states with matching time $t_1 = -4$ are plotted on a semilog plot. The values of the parameters are $m=1$ and $\Lambda=1$.  The solution with no iterations in this case is equivalent to the classical de Sitter solution. From top to bottom, the curves are for the relative differences between the first and zeroth iterations(blue solid), and the first and second iterations(red dashed). }
\label{fig:t1-4-ad-2-compare-iteration-M4}
\end{figure}

We choose $m = 1$ as before and the slightly larger value $\Lambda = 1$ because quantum effects are somewhat
smaller for this state.
The initial time $t_1$ is chosen to be $-4$. This initial condition leads to a bounce solution for this state even though it leads to singular solutions for zeroth order adiabatic states. The scale factors obtained for M4 with two iterations, along with the classical solution are shown in Fig. \ref{fig:t1-4-ad-2-it-2}.  The relative difference between the solutions for the zeroth and first iterations is shown in Fig. \ref{fig:t1-4-ad-2-compare-iteration-M4} along with the relative difference between the first and second iterations.  The relative difference between the first and second iterations is more than an order of magnitude smaller than the relative difference between the zeroth and first iterations.  So it is clear that the iterations are converging.

 \subsection{Fourth-order adiabatic states}
\label{fourth-order-states}

 Solutions to the semiclassical backreaction equations are discussed in this section for
fourth-order adiabatic states.  For the calculations we do, the starting values for the modes are obtained using the method outlined in Appendix~\ref{app:adiabatic-states} and Eqs.~\eqref{W-second-iteration}.

We do not have a good way to combine fourth order adiabatic regularization with the order reduction methods M1, M2, and M3.
However, for the reasons discussed above, it is straight forward to use M4 for fourth order adiabatic vacuum states.

We find that particle production effects are typically smaller for the fourth order states that we use than for the second order or zeroth order adiabatic vacuum states. To illustrate the backreaction effects we therefore choose the value $\Lambda = 1.5$ which is larger than
for our previous illustrations, but we keep $m = 1$ and use an adiabatic matching time of $t_1 = -4$.  In this way, backreaction effects are significant, but not so large as to remove the bounce.

\begin{figure}[htb!]
\centering
\includegraphics[totalheight=0.3\textheight]{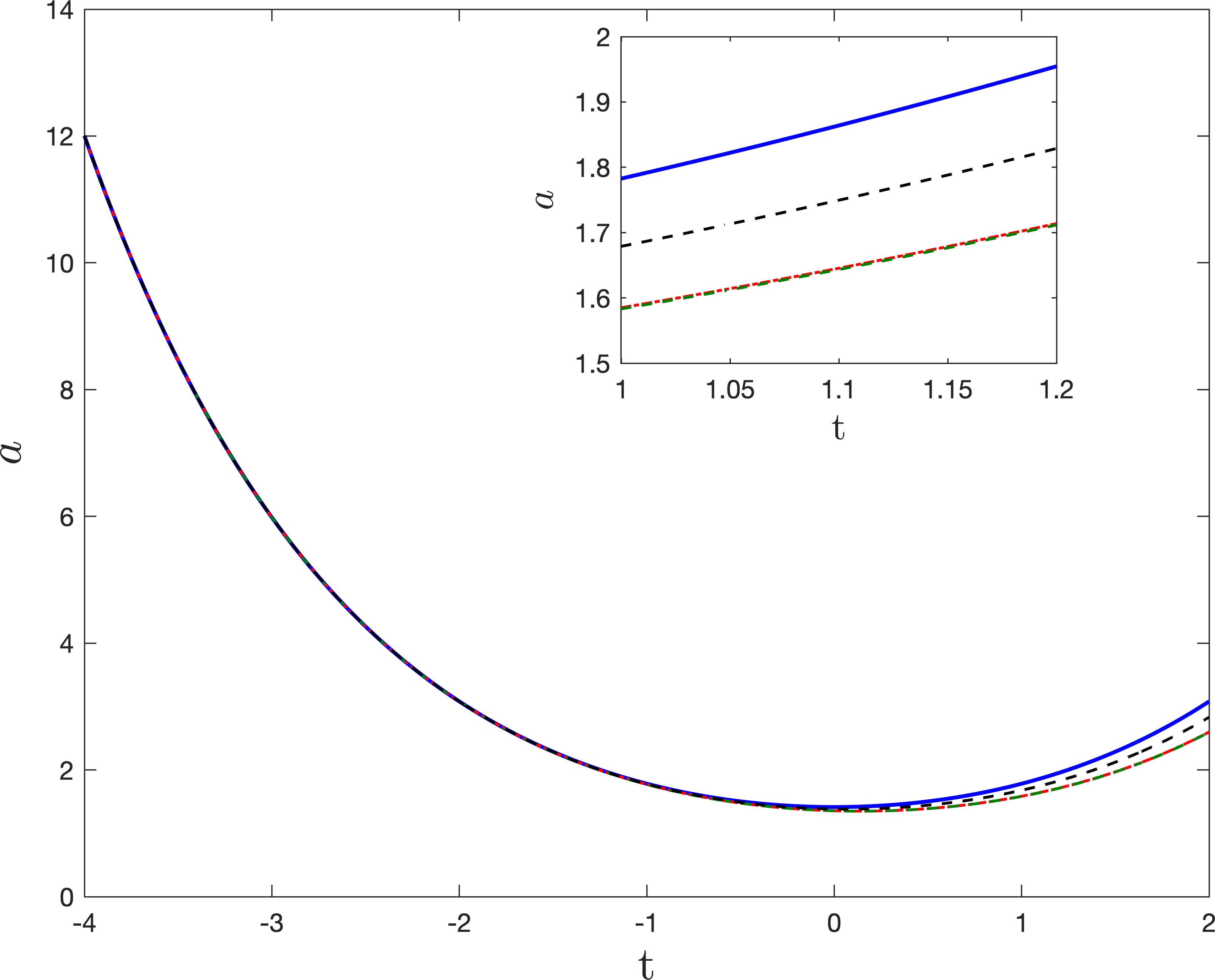}
\caption{Solutions for the scale factor $a$ obtained using M4 for fourth order adiabatic vacuum states with matching time $t_1 = -4$
are plotted.  The values of the parameters are $m=1$ and $\Lambda=1.5$.  Note that the result of the zeroth iteration is the classical de Sitter solution.  From top to bottom, the curves are for the solutions obtained from the zeroth iteration(solid blue), third iteration(dashed black), and first iteration(dash-dotted red). The curve for the second iteration(dashed green) overlaps with the curve for the first iteration. }
\label{fig:t1-4-ad-4-compare-iteration-M4}
\end{figure}

\begin{figure}[htb!]
\centering
\includegraphics[totalheight=0.3\textheight]{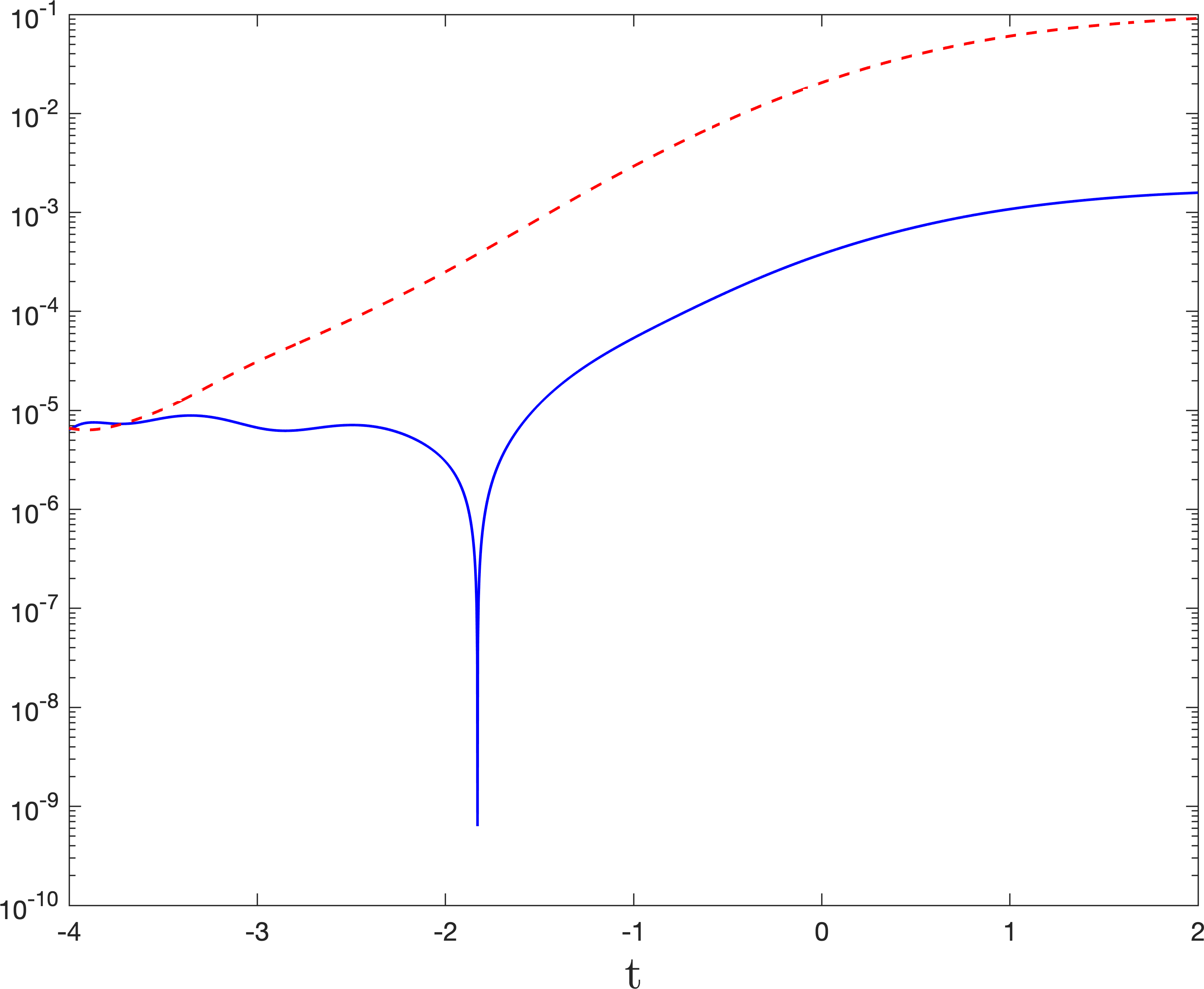}
\caption{Relative differences between the scale factors for some of the results in Fig.~\ref{fig:t1-4-ad-4-compare-iteration-M4} for various iterations using M4 are plotted. 
From top to bottom, the curves are for the relative differences between the second and third iterations(red dashed), and the first and second iterations(blue solid). }
\label{fig:t1-4-ad-4-compare-relative-iteration-M4}
\end{figure}

As discussed in Sec.~\ref{sec:order-reduction}, M4 involves first solving the classical Einstein equations, using that solution to compute the stress-energy tensor for the quantum field, and then solving the semiclassical backreaction equations using that stress-energy tensor as a source. This is the first iteration.  For the second iteration, the solution from the first iteration is used to compute the stress-energy tensor for the quantum field, and so forth.
The classical solution and the results for the first three iterations are shown in Fig. \ref{fig:t1-4-ad-4-compare-iteration-M4}.
The relative differences between the first and second iterations and between the second and third iterations are shown in Fig.~\ref{fig:t1-4-ad-4-compare-relative-iteration-M4}.

It is clear from the figures that there is only a relatively small difference between the first and second iterations.  Thus the iterations appear to be converging as expected.  However, the difference between the second and third iterations is significantly larger.

After a thorough investigation, we find that one significant difference between the second and third iterations is the accuracy of the fits used to compute $\dddot{a}$ and $\ddddot{a}$ using the scale factor obtained from the previous iteration.  The fits are then used to compute $\la T_{ab} \ra$ and to compute the starting values for the mode functions which are discussed in Appendix~\ref{app:adiabatic-states}.  Since there was no problem with the convergence of iterations for zeroth and second order adiabatic states, we strongly suspect
that the primary effect of the poorer fits for the third iteration is in computing the starting values for the modes.  These do not contain third and fourth derivatives of the scale factor for zeroth and second order adiabatic states.  The errors in the fits effectively change the state of the quantum field.  This in turn results in a different solution to the backreaction equations than would be obtained with a more accurate fit.

To make the fits, we first fit the numerical data for $\ddot{a}$ that comes from the previous iteration.  Then derivatives of that fit are used to obtain fits for $\dddot{a}$ and $\ddddot{a}$.  These derivatives almost certainly result in less accuracy for these fits than occurs for the fit for $\ddot{a}$.  This loss of accuracy should affect the accuracy of the solutions obtained by the next iteration.  In this way, the accuracy of the results should decrease for each iteration.
Of course in principle one can simply increase the accuracy of all of the calculations to allow for more iterations.  However, there is still expected to be an increased error with each iteration and eventually this will become significant.

\begin{figure}[htb!]
\centering
\includegraphics[totalheight=0.3\textheight]{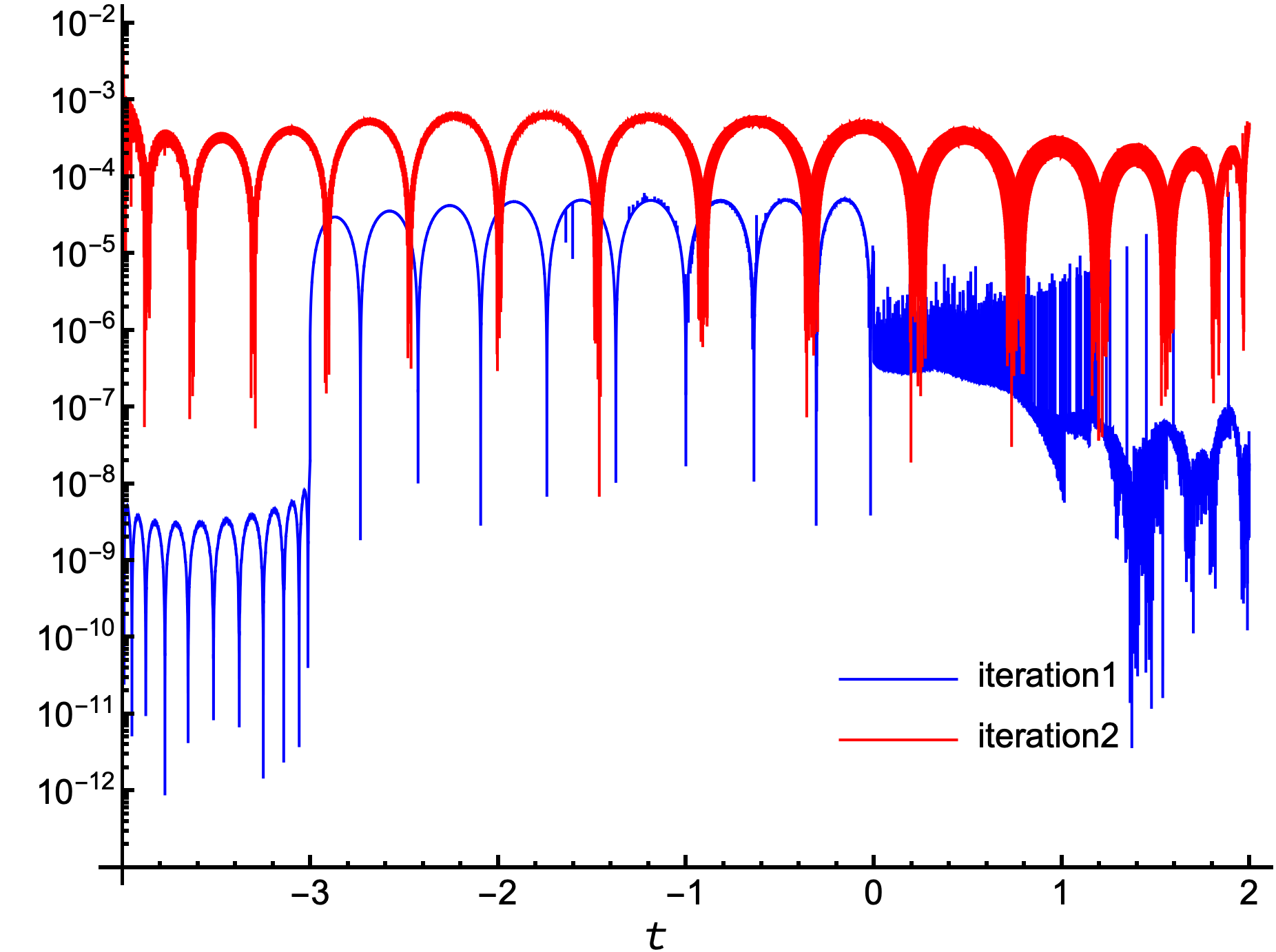}
\caption{The relative differences between the fits for $\ddot{a}$ and the numerical data used to make the fits for the numerical computations of the first and second iterations that are shown in Fig.~\ref{fig:t1-4-ad-4-compare-iteration-M4}.  The lower (blue) curve corresponds to the relative difference for the first iteration and the upper (red) curve corresponds to the relative difference for the second iteration.  Clearly there is significantly less accuracy for the fit using the second iteration results  than for the fit using the first iteration results.   }
\label{fig:t1-4-ad-4-compare-fit-iteration-M4}
\end{figure}

The evidence that the fits are less accurate for successive iterations comes from comparing each fit for $\ddot{a}$ with the numerical data used to make the fit.  In Fig.~\ref{fig:t1-4-ad-4-compare-fit-iteration-M4}, the relative differences between the fit and the data are plotted for the fits made after the first and second iterations.  It is clear that the accuracy of the fit for the first iteration is significantly better than the accuracy for the second iteration.  In each case, the fits were obtained by fitting the data to a power series in the time coordinate $t$.  Terms were added to the power series until it was found that adding more terms did not increase the accuracy of the fit.
Further, when $\ddot{a}$ for the first iteration was fitted, the numerical data was separated into several overlapping segments to improve the fit accuracy. However, this strategy did not work for the second iteration and even made the accuracy worse, so the fit shown is for the entire data set.

\section{Particle production}
\label{sec:particle-production}

In general in a curved spacetime it is not possible to uniquely separate out particle production from vacuum polarization effects.
However, there is an important exception: there is no particle production for conformally invariant quantum fields in
conformally flat spacetimes.  In Sec.~\ref{sec:no-class-rad}, the semiclassical backreaction equations were solved when only conformally invariant
quantum fields in the conformal vacuum state are present and the cosmological constant is positive. It was found that the order reduction methods resulted in the spacetime remaining de Sitter space (to the level of approximation appropriate for each method) but vacuum polarization effects alter the effective cosmological constant.
To take into account in a classical way the possible effects of particle production, classical radiation was included along with conformally
invariant fields in Sec.~\ref{sec:class-rad}.  It was found that if enough radiation is present, then collapse to a final singularity occurs for solutions to the semiclassical backreaction equations for each of the order reduction methods.  Otherwise a bounce occurs.

In Sec.~\ref{sec:massive-field}, the semiclassical backreaction equations were solved when a massive conformally coupled scalar field is present.  Here one cannot uniquely separate particle production from vacuum polarization effects.  However, examination of~\eqref{Tab-renorm-method} and~\eqref{Tab-analytic} shows that the stress-energy tensor can be divided into a part that is identical to the stress-energy tensor for a massless conformally coupled scalar field and a second part that consists of a sum over products of the mode functions and their derivatives along with certain subtraction terms that do not contain higher derivatives of the scale factor.  Thus, we can write
\bea \la T_{ab} \ra_r &=& \la T_{ab} \ra_{\rm massive} + \la T_{ab} \ra_{m = 0} \\
\la T_{ab} \ra_{\rm massive} &=& \la T_{ab} \ra_n + (\la T_{ab} \ra_{\rm an} -  \la T_{ab} \ra_{m = 0}) \nonumber \\
 \la T_{ab} \ra_{m = 0} &=& \frac{1}{2880 \pi^2} \left[ -\frac{1}{6}\; ^{(1)}H_{ab} +  ^{(3)}H_{ab} \right] \;. \nonumber \eea

It is clear that the particle production must come from $\la T_{ab} \ra_{\rm massive}$. However, this term may also contain vacuum polarization effects.  To investigate how important the vacuum polarization effects are for this term and for the stress-energy tensor in general we consider two solutions to the semiclassical backreaction equations and focus on the energy density of the field
$ \la \rho \ra = \la T_{tt} \ra $

\begin{figure}[htb!]
\centering
\includegraphics[totalheight=0.3\textheight]{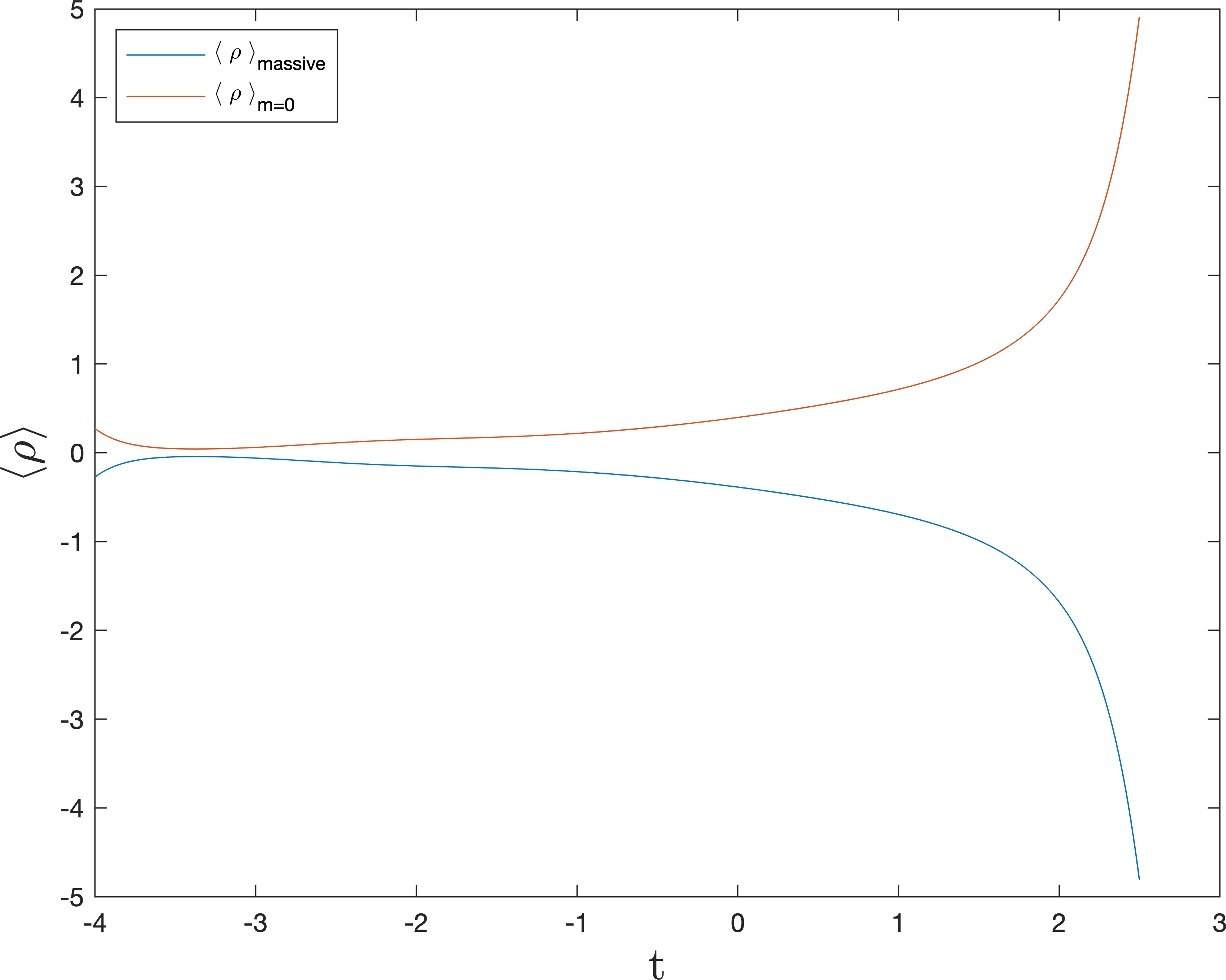}
\caption{Solutions for the energy density $\langle \rho \rangle$ plotted versus the proper time $t$ using M4. A zeroth-order adiabatic vacuum state is used with matching time $t_1=-4$ with two iterations. The values of the parameters are $m=1$, $\Lambda=0.5$. From top to bottom, the curves are for $\langle \rho \rangle_{m=0}$ (red) and $\langle \rho \rangle_{\mbox{massive}}$ (blue). }
\label{fig:t1-4-zeroth-adiabatic-M4-rho}
\end{figure}

\begin{figure}[htb!]
\centering
\includegraphics[totalheight=0.3\textheight]{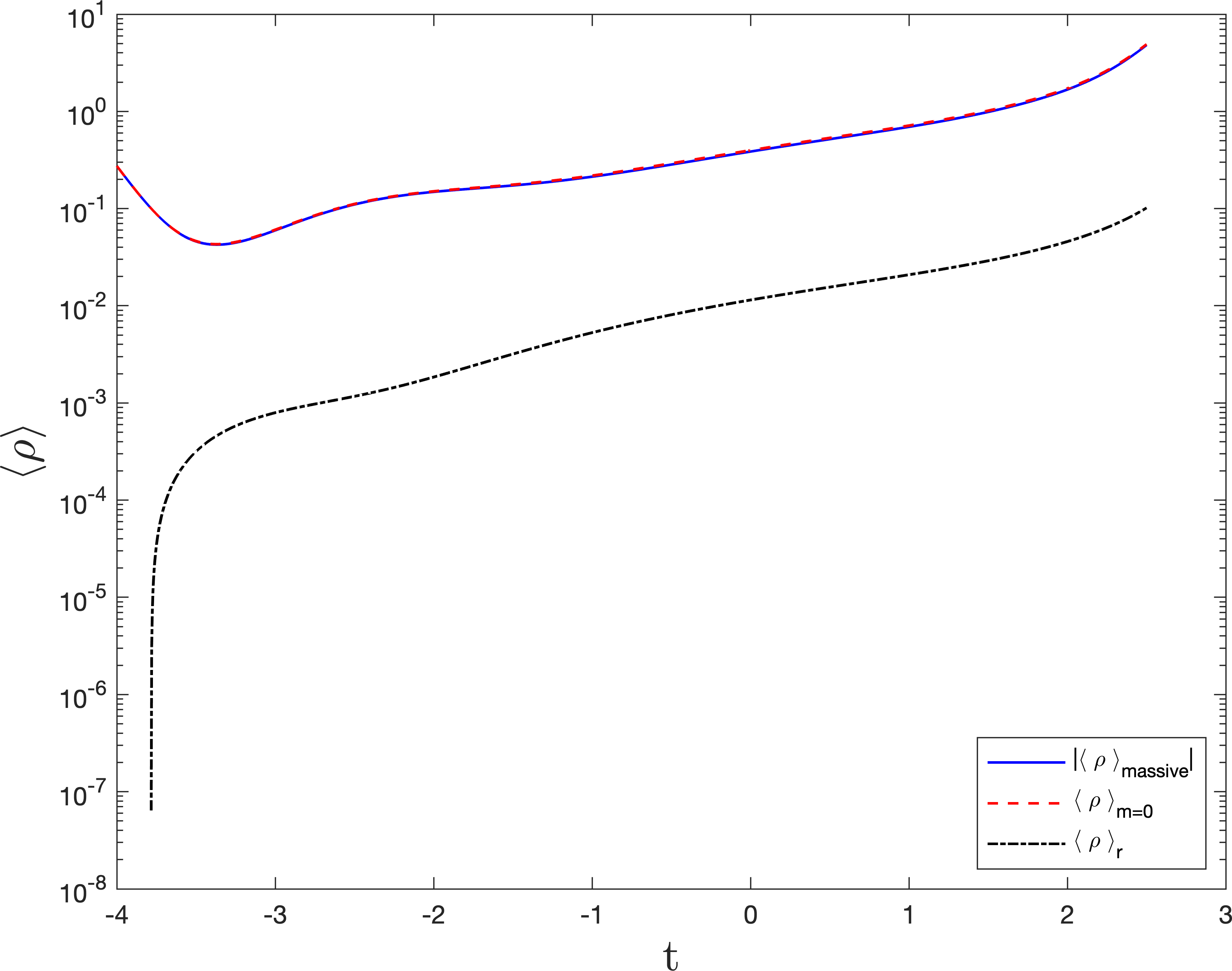}
\caption{Solutions for the energy density $\langle \rho \rangle$ plotted versus the proper time $t$ using M4 on a semilog plot. Zeroth-order adiabatic vacuum states are used with matching time $t_1=-4$ with two iterations. The values of the parameters are $m=1$, $\Lambda=0.5$. From top to bottom, the curves are for $\langle \rho \rangle_{m=0}$ (dashed red), the absolute value of $\langle \rho \rangle_{\mbox{massive}}$ (solid blue), and $\langle \rho \rangle_r$ (dash-dotted black). Note that the curves for $\langle \rho \rangle_{m=0}$ and $|\langle \rho \rangle_{\mbox{massive}}|$ overlap. }
\label{fig:t1-4-zeroth-adiabatic-M4-rho-semilog}
\end{figure}

The first solution we consider is the one discussed in Sec.~\ref{zeroth-order-ad} for a zeroth order adiabatic vacuum state that begins at time $t_1 = -4$ and ends in a final singularity.  For this solution $m =1$ and $\Lambda = 0.5$.  We consider the specific solution that was computed with two iterations using M4.  Because the full stress-energy tensor is computed in the geometry of the previous iteration for M4, the energy density shown is that computed with the scale factor obtained from the first iteration.  Both $\la \rho \ra_{\rm massive}$ and $\la \rho \ra_{m=0}$ are shown in Fig.~\ref{fig:t1-4-zeroth-adiabatic-M4-rho}.  It is clear that vacuum polarization effects make a significant contribution to $\la \rho \ra_{\rm massive}$ since there is a partial cancelation of this term by $\la \rho \ra_{m=0}$ when the two are added together.  This is shown explicitly in Fig.~\ref{fig:t1-4-zeroth-adiabatic-M4-rho-semilog} where the absolute values of these contributions to the energy density are plotted along with the total energy density.

To understand why there is so much cancelation when $\la \rho \ra_{\rm massive}$ and $\la \rho \ra_{m=0}$ are added together, consider the WKB approximation.  It is the WKB approximation~\eqref{WKB-psi} that is used to compute the renormalization counterterms $\la T_{ab} \ra_{\rm ad}$ in~\eqref{Tab-r-first}~\cite{anderson-parker}. A fourth order approximation is used to do this.  The mode functions in $\la T_{ab} \ra_u$ are computed numerically without the use of the WKB approximation (except for their starting values).  The strong cancelation indicates that, over the range of times in the plot, the WKB approximation must be a good approximation for the exact modes.  Examination of Fig.~\ref{fig:t1-4-ad-0-it-2} shows that the scale factor is approaching the final singularity during the times shown.  One would guess that the WKB approximation would break down in the final stages of approach to the singularity due to the rapid change in the scale factor with time.  However, it is difficult to carry out the iterations as the singularity is approached because of the increase in the rate of change of the scale factor makes it more difficult to fit $\ddot{a}$ accurately enough to obtain accurate values for the higher derivatives in $\la T_{ab} \ra_{m = 0}$.

\begin{figure}[htb!]
\centering
\includegraphics[totalheight=0.3\textheight]{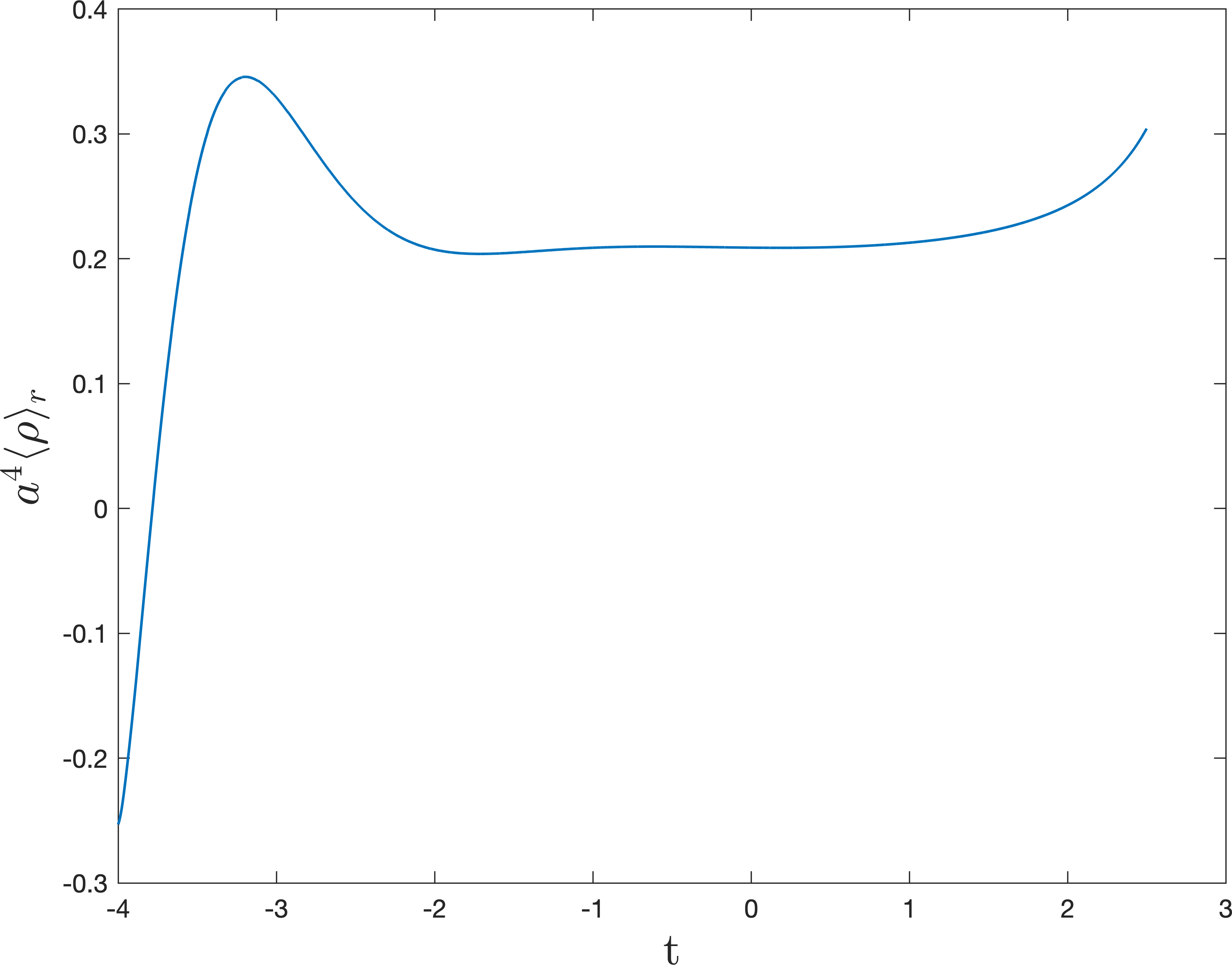}
\caption{Solutions for the energy density $a^4\langle \rho \rangle_r$ plotted versus the proper time $t$ using M4. Zeroth-order adiabatic vacuum states are used with matching time $t_1=-4$ with two iterations. The values of the parameters are $m=1$, $\Lambda=0.5$. }
\label{fig:t1-4-zeroth-adiabatic-M4-rhoa4}
\end{figure}

Since there is so much cancelation between $\la \rho \ra_{\rm massive}$ and $\la \rho \ra_{m=0}$ in this case, we use the total energy density to
investigate how much of a contribution the particles make to the stress-energy tensor.  Because the particles are massive, we would expect their energy density to go like $a^{-3}$ when the scale factor is large.  As can be seen from the mode equation~\eqref{mode-eq-1}, they are effectively massless when the scale factor is small and we would thus expect their energy density to go like $a^{-4}$ in this case.  For the range of times that we consider the particles only provide the dominant contribution to $\la \rho \ra_r$ when they are effectively massless.  In Fig.~\ref{fig:t1-4-zeroth-adiabatic-M4-rhoa4},  the quantity $a^4 \la \rho \ra_r$ is plotted.  One can see that there is a range of times for which this is a constant, which indicates that it is dominated by particles that are effectively massless.  However, at the latest times in the plot, where the scale factor is smallest, this quantity increases which indicates that the energy density is growing faster than $a^{-4}$.  Thus vacuum polarization effects are dominating near the final singularity, as might be expected.

\begin{figure}[htb!]
\centering
\includegraphics[totalheight=0.3\textheight]{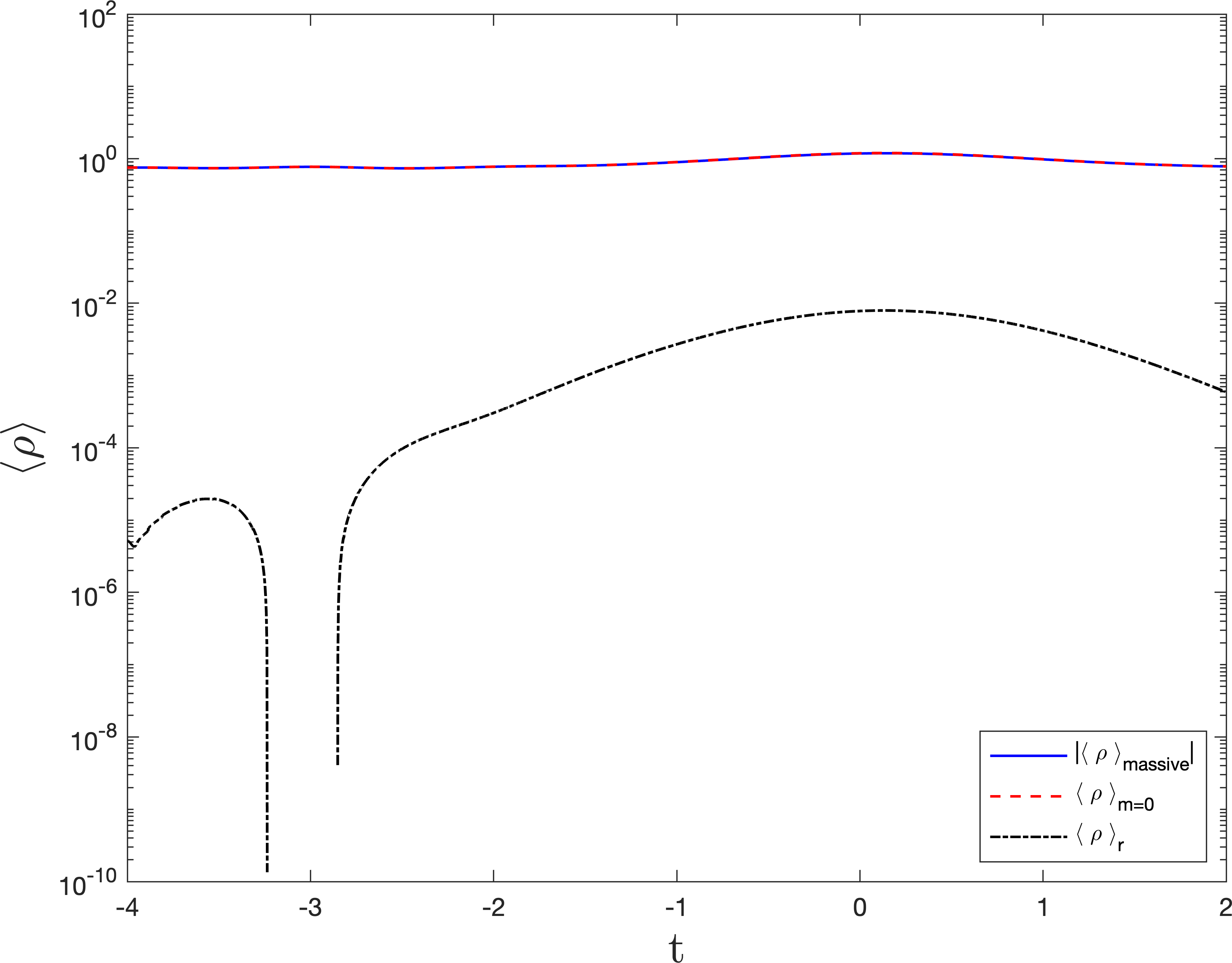}
\caption{Solutions for the energy density $\langle \rho \rangle$ plotted versus the proper time $t$ using M4 on a semilog plot. Fourth-order adiabatic vacuum states are used with matching time $t_1=-4$ with two iterations. The values of the parameters are $m=1$, $\Lambda=1.5$. From top to bottom, the curves are for $\langle \rho \rangle_{m=0}$ (dashed red), $|\langle \rho \rangle_{\mbox{massive}}|$ (solid blue), and $\langle \rho \rangle_r$ (dash-dotted black). Note that the $\langle \rho \rangle_{m=0}$ and $|\langle \rho \rangle_{\mbox{massive}}|$
 overlap. }
\label{fig:t1-4-fourth-adiabatic-M4-rho-semilog}
\end{figure}

\begin{figure}[htb!]
\centering
\includegraphics[totalheight=0.3\textheight]{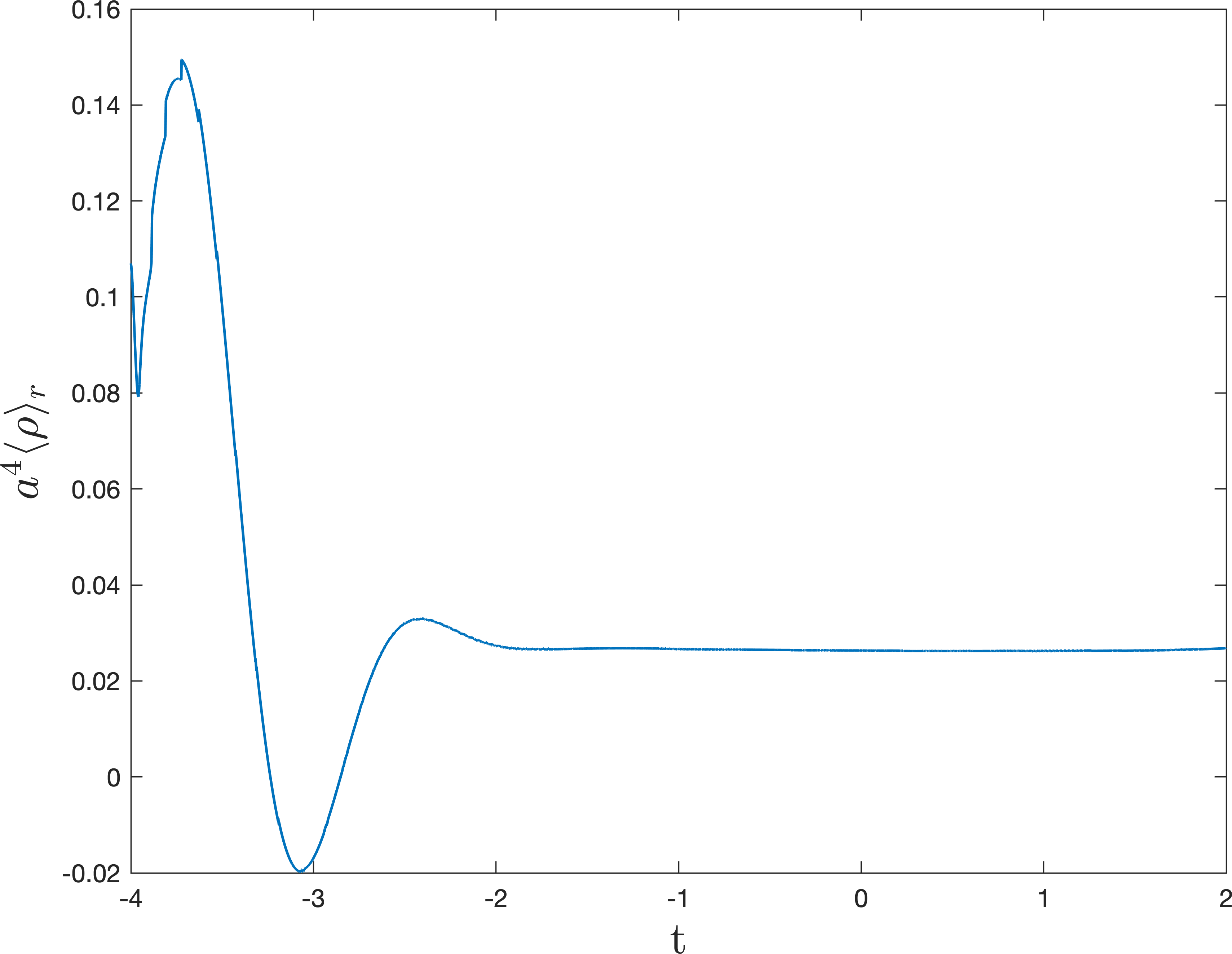}
\caption{Solutions for the energy density $a^4\langle \rho \rangle_r$ plotted versus the proper time $t$ using M4. Fourth-order adiabatic vacuum states are used with matching time $t_1=-4$ with two iterations. The values of the parameters are $m=1$, $\Lambda=1.5$. }
\label{fig:t1-4-fourth-adiabatic-M4-rhoa4}
\end{figure}

The second solution we consider is discussed in Sec.~\ref{fourth-order-states}.  It is for a fourth order adiabatic state with $m = 1$ and $\Lambda = 1.5$.  It  begins at $t_1 = -4$ and undergoes a bounce.  M4 is used and, as in the previous example, $\la \rho \ra_r$ is computed using the scale factor that results from the first iteration.  As in the previous example, it is found for the range of times investigated, that there is a significant cancelation between $\la \rho \ra_{\rm massive}$ and $\la \rho \ra_{m=0}$ when they are added together.  This is illustrated in Fig.~\ref{fig:t1-4-fourth-adiabatic-M4-rho-semilog} where the absolute values of these quantities along with that of $\la \rho \ra_r$ are plotted.  As for the previous example, the particles never dominate $\la \rho \ra_r$ when they act like massive particles for the times considered.  However, they provide the dominant contribution to $\la \rho \ra_r$ near the bounce, where they are acting like massless particles as is shown in Fig.~\ref{fig:t1-4-fourth-adiabatic-M4-rhoa4}.

In~\cite{am-particle}, the effects of particle production due to a massive scalar field in de Sitter space in spatially closed coordinates were investigated in the test field approximation where backreaction effects due to the quantum field are not taken into account.  It is shown in Fig. 11 of that paper that near the bounce the particles for $m = H = \sqrt{\frac{\Lambda}{3}}$ dominated the energy density for a second order adiabatic state.  In Fig 12 it is shown that they are effectively massless near the bounce.  Our results in Fig.~\ref{fig:t1-4-fourth-adiabatic-M4-rhoa4} are consistent with those results.

\section{Summary}
\label{sec:summary}

We have investigated backreaction effects due to both conformally invariant fields and conformally coupled massless scalar fields in closed Roberson-Walker spacetimes when a positive cosmological constant is present.  It is assumed that the fields are in homogeneous and isotropic states so backreaction effects do not destroy the homogeneity and isotropy. In one case classical radiation is also included.  Because the spacetimes are conformally flat no particle production occurs for the conformally invariant fields.  In contrast, the conformally coupled massive scalar field is not conformally invariant and so particle production can occur.  For this field we consider adiabatic vacuum states of various orders instead of the Bunch-Davies state.  As argued in~\cite{am-particle} there is a sense in which these are more natural states for this field in a spacetime that begins like collapsing de Sitter space.

The stress-energy tensors for the quantum fields contain higher derivative terms.  For the massive conformally coupled scalar field these are exactly the same as for the massless field.  None of the terms which involve the mass contain higher derivatives of the scale factor.  We have used four different methods to reduce the resulting fourth order semiclassical backreaction equations to second order.  The method we call M1 is a direct adaptation of the Parker-Simon method to the case when particle production occurs.  M2 involves solving the semiclassical backreaction equations when the terms proportional to $^{(1)}H_{ab}$ and $^{(3)}H_{ab}$ in $\la T_{ab} \ra$ are ignored for the zeroth iteration.
These terms are evaluated in the resulting metric and used as source terms for the next iteration.  Further iterations proceed in the same manner.  M3 involves solving the semiclassical equations when only the terms proportional to $^{(1)}H_{ab}$ are ignored for the zeroth iteration.  These are the terms that contain third and fourth derivatives of the scale factor.  The iterations work in the same way as for the second method.  M4 is due to Agullo~\cite{agullo}.  It involves starting with the classical solution, evaluating the full stress-energy tensor for the quantum fields in that geometry, and using it as a source term for the next iteration and so forth.

If no classical matter is present and quantum effects are ignored, then the solution to Einstein's equations is the de Sitter solution.  It undergoes a time-symmetric bounce at a value of the scale factor that is inversely proportional to the square root of the cosmological constant $\Lambda$.  If classical radiation is present then the bounce occurs at a smaller value of the scale factor or, if enough radiation is present, the Universe collapses to a singularity.  This same effect is  found for the massive scalar field when no classical radiation is present.  If enough particle production occurs then the bounce is removed and appears to be replaced with a singularity.

For the case when there is no classical radiation and only conformally invariant quantized fields in the conformal vacuum state, it turns that all of the methods result in exact de Sitter space solutions with various values of the parameter $H = \sqrt{\Lambda_{\rm eff}/3}$, with $\Lambda_{\rm eff}$ the effective cosmological constant.  For the exact solution to the semiclassical backreaction equations, $H$ depends on the value of the actual cosmological constant $\Lambda$ and the parameter $\beta_q$ in~\eqref{betaq}.  Because the term proportional to $\alpha_q$ in~\eqref{alphaq} vanishes for the de Sitter metric, M3 is equivalent to solving the exact semiclassical backreaction equations.  If $H^2$ is expanded in powers of $\beta_q$, then M1 results in an expression for $H^2$. that is equal to the zeroth and first order terms in that expansion.  M2 and M4 are equivalent in this case.  The i'th iteration reproduces the expansion of the exact solution to order $(\beta_q)^i$, but it is not exactly the same because there are terms proportional to higher powers of $\beta_q$.

If classical radiation is present, then exact de Sitter space is no longer a solution to the classical Einstein equations.  The various possibilities are discussed in Appendix~\ref{app:classical}.  For this paper the focus is on classical solutions which undergo a bounce.  These will occur if there is not too much classical radiation present.  When only conformally invariant quantum fields are present the different methods give slightly different modifications of the classical solution with bounces that occur at smaller values of the scale factor than the classical solution for the examples investigated.  In this case M2 and M4 are equivalent.  It is found that near the bounce only small differences occur between the classical solution and the solutions obtained by each method.  M2 and M3 involve iterations, and in the examples investigated, it was found that the iterations appear to converge.  However, the scale factors that the iterations appear to converge to are slightly different for M2 than M3.

The semiclassical backreaction equations were also solved when a conformally coupled massive scalar field was present but there was no classical radiation and no conformally invariant fields.  For this field it turns out that it is mathematically consistent in a homogeneous and isotropic spacetime to consider zeroth and second order adiabatic states as well as the fourth order adiabatic states that work for arbitrarily coupled scalar fields.  All four of the methods we use work for zeroth order adiabatic states.  The issue for higher order adiabatic states is how to fix the initial conditions.  We were not able to adapt the Parker-Simon approach to second or fourth order adiabatic states.  M2, M3, and M4 can also be used for second order adiabatic states.  However, only M4 works for fourth order and higher adiabatic states.

Because particle production occurs for the massive scalar field, if one starts with a solution that is initially contracting, there are two possibilities.  One is that, as for the corresponding classical solution, the cosmological constant causes a bounce to occur.  The other is that enough particle production occurs that there is no bounce and the universe contracts to a final singularity.  As for the case when conformally invariant fields are present and no particle production occurs, the different methods result in slightly different solutions.  In principle there could be cases in which one method results in a bounce and another in contraction to a final singularity, but these would only occur for fine tuning of the initial conditions.  For the cases considered here, all methods either gave a bounce or contraction to a singularity.  In both cases, the solutions for the different methods tend to get farther apart as time progresses.  However, this happens much more rapidly for the solutions which approach a final singularity.

We also examined the behavior of the energy density of the massive scalar field for a solution to the semiclassical backreaction equations that approaches a final singularity at $a = 0$ and one that undergoes a bounce.  In both cases we find that for the range of times considered, the WKB approximation for the modes of the quantum field is a relatively good approximation.  We come to this conclusion because there are cancelations between the part of the energy density that depends directly on the modes and the part that survives in the massless limit and results entirely from a fourth order WKB approximation for the modes.  Particle production is a nonlocal effect and cannot be obtained directly from the type of WKB approximation we use here.  Our results show that after the cancelations, the full energy density of the quantum field is dominated by the particles for a range of times as the singularity is approached, but vacuum polarization effects become important at later times.  The particles at these times are effectively massless.  For the solution undergoing a bounce it was found that the energy density is dominated by particles near the bounce.  Again the particles during this time period are effectively massless.  The results for the second solution are consistent with those found in~\cite{am-particle} for a massive field in de Sitter space when semiclassical backreaction effects are not taken into account.

Given that the four methods of order reduction result in very similar solutions to the semiclassical backreaction equations, one can ask which method or methods is preferred.  For calculations that involve only conformally invariant quantum fields, M1 is the easiest to use because it involves no iteration.  If quantum effects are very small then this is probably the preferred method.  If quantum effects are more significant, then one of the methods involving iteration may give a more accurate answer.  If conformally noninvariant quantum fields are present then M4 is the preferred method because it is straightforward to implement it for a state of any adiabatic order.  We have shown that it gives similar answers to the other methods for zeroth order adiabatic states for the massive conformally coupled scalar field.  Thus, there is good reason to believe that it will be a reliable method of order reduction for more general cases including spacetime geometries that are not conformally flat and/or quantum fields that are not conformally invariant.

\acknowledgments

We would like to thank Ivan Agullo and Eric Carlson for helpful conversations, and Emil Mottola for helpful comments on the manuscript.  PRA would like to thank the Northwestern's Center for Interdisciplinary Exploration and Research in Astrophysics (CIERA) for hosting him during a visit in August of 2023.  This work was supported in part by the National Science Foundation under Grants No. PHY-1505875, PHY-1912584, and PHY-2309186  to Wake Forest University.  Some of the numerical work was done using the WFU DEAC Cluster;  we thank the WFU Provost's Office and Information Systems Department for their generous support.

\appendix

\section{WKB Approximation and Adiabatic Vacuum States}
\label{app:adiabatic-states}

The states we consider for the massive scalar field are adiabatic vacuum states~\cite{b-d-book}.  We construct them using the WKB approximation for the modes.  A zeroth order adiabatic state is constructed using a zeroth order WKB approximation, a second order adiabatic state is constructed using a
second order adiabatic approximation, and so forth.

There are two common time coordinates used in cosmology.  One is the proper time $t$ which we use here when solving the semiclassical backreaction
equations.  The natural form for the time dependent part of the mode function in this case is $f_k$.  The other is the conformal time $\eta$ for which the natural time dependent part of the mode function is $\psi_k$.  The relation between the times is $dt = a d \eta$ and that between the mode functions is $f_k = \sqrt{a} \, \psi_k$.  One can derive WKB approximations for both $\psi_k$ and $f_k$ using the same type of ansatz.  For $\psi_k$ one has
\bes \bea \psi_k &=& \frac{\exp(-i \int_{\eta_0}^\eta W(x) dx)}{\sqrt{2 W}} \label{WKB-psi-eq} \\
 W^2 &=& \Omega^2 - \frac{W''}{2 W} + \frac{3}{4} \frac{W'^2}{W^2} \label{WKB-exact-eq} \\
     \Omega^2 &=& k^2 + m^2 a^2 \label{Omega-def}
  \eea \label{WKB-psi} \ees
  with $\eta_0$ an arbitrary real number.
This can be solved by iteration with the zeroth order term being
\be W_k^{(0)} = \Omega \label{W-zeorth-iteration} \;. \ee
For $f_k$, the WKB expansion in terms of proper time is
\bes \bea f_k &=& \frac{\exp(-i \int_{t_0}^t \bar{W}(x) dx)}{\sqrt{2 \bar{W}}} \;, \\
 \bar{W}^2 &=& \bar{\Omega}^2 - \frac{\ddot{\bar{W}}}{2 \bar{W}} + \frac{3}{4} \frac{\dot{\bar{W}}^2}{\bar{W}^2} \;, \\
 \bar{\Omega}^2 &=& \frac{k^2}{a^2} + m^2 + \frac{\dot{a}^2}{4 a^2} - \frac{1}{2} \frac{\ddot{a}}{a} \; ,  \eea \label{WKB-fk} \ees
where $t_0$ is an arbitrary real number.

One can fix an adiabatic vacuum state by choosing a matching time $\eta_1$ and using a particular order of the WKB
approximation to obtain starting values for $\psi_k$ and $\psi^{'}_k$.  It is convenient to choose $\eta_0 = \eta_1$ in
which case the zeroth order adiabatic approximation gives
\be \psi_k(\eta_1) = \frac{1}{\sqrt{2 \Omega(\eta_1)}} \;. \label{psi-ad-0} \ee
There is an ambiguity in the first derivative.  One can either choose
\bes \be \psi^{'}_k(\eta_1) = -i \sqrt{\frac{\Omega}{2}} \;, \label{psiprime-ad-0-1}\ee
or
\be \psi^{'}_k(\eta_1) = -i \sqrt{\frac{\Omega}{2}} - \frac{\Omega^{'}}{(2 \Omega)^{3/2}} \;. \label{psiprime-ad-0-2} \ee \ees
Substitution of~\eqref{psi-ad-0} and either~\eqref{psiprime-ad-0-1} or~\eqref{psiprime-ad-0-2} into~\eqref{Tab-eta} and subtracting off
the renormalization counterterms results in finite expressions for the energy density and trace.  Thus both expressions result in acceptable zeroth order adiabatic states.

It is worth emphasizing that for any adiabatic state the solutions to the mode equation are exact solutions.  The WKB approximation is
only used to obtain starting values for these solutions at some time $\eta_0$.

Since there is a WKB approximation for $f_k$, one would guess that one could also use the WKB expansion~\eqref{WKB-fk} to define adiabatic states.
While this is correct, it is not as straightforward as it is when using~\eqref{WKB-psi}.  To see this, one can try following the same procedure as above but using~\eqref{WKB-fk}.  Then,
\bea f_k(t_1) &=& \frac{1}{\sqrt{2 \bar{\Omega}}} \nonumber  \\
     \dot{f}_k(t_1) &=& -i \sqrt{\frac{\bar{\Omega}}{2}}  \label{f-ad-bad}
\eea
The problem is that if~\eqref{f-ad-bad} is substituted into~\eqref{Tab-t} and the adiabatic counterterms are subtracted, one
finds that all of the ultraviolet divergences in $\la \rho \ra_{\rm u}$ have not been cancelled.

Because of this we use the following method to obtain an adiabatic vacuum state of order $n$.  First iterate~\eqref{WKB-exact-eq} $\frac{n}{2}$ times to obtain the WKB approximation to the correct order.  We will call the solution $W_k^{(n)}$.  Then substitute it into~\eqref{WKB-psi-eq} with $\eta_0 = \eta_1$ and evaluate at time $\eta = \eta_1$.  Next compute the time derivative of~\eqref{WKB-psi-eq} and substitute the WKB expansion into it, keeping only terms up to and including the terms with $n$ derivatives of the scale factor.  The result is
\bes \bea \psi_k(\eta_1) &=& \sqrt{\frac{1}{2W_k^{(n)}(\eta_1)}} \;, \\
     \psi_k^{'}(\eta_1) &=& -\frac{W_k^{(n)\;'}(\eta_1)}{\left(2 W_k^{(n)}(\eta_1) \right)^{3/2}}-i\sqrt{\frac{W_k^{(n)}(\eta_1)}{2}} \;, \\
 \eea \label{psi-initial-wkb} \ees
For zeroth order adiabatic states we use~\eqref{W-zeorth-iteration} and $W_k^{(0)\;'} = 0$.
For second order  adiabatic  states we use
\bes \bea W_k^{(2)} &=& W_k^{(0)}+\frac{5m^4a^2a'^2}{8 {W_k^{(0)}}^5}-\frac{m^2a'^2}{4 {W_k^{(0)}}^3}-\frac{m^2aa''}{4 {W_k^{(0)}}^3} \;, \\
    W_k'^{(2)}&=&\frac{m^2aa'}{W_k^{(0)}} \;,  \eea
\label{W-first-iteration} \ees
and for fourth order adiabatic states we use
 \bes \bea   W_k^{(4)}&=& W_k^{(0)}+\frac{5m^4a^2a'^2}{8 {W_k^{(0)}}^5}-\frac{m^2a'^2}{4 {W_k^{(0)}}^3}-\frac{m^2aa''}{4 {W_k^{(0)}}^3}-\frac{1105m^8a^4a'^4}{128{W_k^{(0)}}^{11}} \nonumber \\
     & & \; +\frac{221m^6a^2a'^4}{32{W_k^{(0)}}^9}-\frac{19m^4a'^4}{32{W_k^{(0)}}^7}
    +\frac{221m^6a^3a'^2a''}{32{W_k^{(0)}}^9}-\frac{61m^4aa'^2a''}{16{W_k^{(0)}}^7}  \\
    && \; -\frac{19m^4a^2a''^2}{32{W_k^{(0)}}^7}+\frac{3m^2a''^2}{16{W_k^{(0)}}^5}-\frac{7m^4a^2a'a'''}{8{W_k^{(0)}}^7}
    + \frac{m^2a'a'''}{4{W_k^{(0)}}^5}+\frac{m^2aa''''}{16{W_k^{(0)}}^5} \;,
   \nonumber \\
     W_k'^{(4)}&=&\frac{m^2aa'}{W_k^{(0)}}-\frac{25m^6a^3a'^3}{8{W_k^{(0)}}^7}+\frac{2m^4aa'^3}{{W_k^{(0)}}^5}
     +\frac{2m^4a^2a'a''}{{W_k^{(0)}}^5}-\frac{3m^2a'a''}{4{W_k^{(0)}}^3}-\frac{m^2aa'''}{4{W_k^{(0)}}^3} \;. \nonumber \\
  \eea \label{W-second-iteration} \ees
     These can be converted to starting values for $f_k$ and $\dot{f}_k$ by substituting the appropriate expressions for a given adiabatic order into~\eqref{psi-initial-wkb} and using the relations $dt = a d \eta$ and $ f_k = a^{1/2} \psi_k$.

\section{Solutions to the classical Einstein equations when radiation and a cosmological constant are present}
\label{app:classical}

 If we set both $\la 0|\rho |0 \ra$ in~\eqref{rho-eq} and $\la 0 |T | 0\ra$ in~\eqref{T-eq} to zero, then the classical Einstein equations are
\bes \bea \alpha^{'\; 2} &=& \frac{A}{\alpha^2} - 1 + \frac{\alpha^2}{3} \;, \label{alp-eq1} \\
          \alpha^{''} &=& -\frac{A}{\alpha^3} + \frac{\alpha}{3} \;, \label{alp-eq2} \eea \label{alp-eqs} \ees
where~\eqref{rho-eq} has been used to obtain the second equation.
The solution that gives the Einstein universe occurs for $A = \frac{3}{4}$ and is
\be \alpha = \sqrt{\frac{3}{2}}  \;. \label{Einstein-univ} \ee
Two exact dynamical solutions for $A = \frac{3}{4}$ are
\bea \alpha_{\pm} &=& \sqrt{-A e^{\pm 2 T} + 3 \cosh^2 T } \;, \label{alp-sol} \\
     T &\equiv& \frac{\tau - \tau_0}{\sqrt{3}} \;, \nonumber \eea
     with $\tau_0$ an arbitrary constant.
Note that $\alpha_{\pm}$ both reduce to the de Sitter solution $\alpha = \sqrt{3} \cosh T$ in the limit $A \to 0$.

For $A>0$ there are three cases that need to be considered separately.
For $A = \frac{3}{4}$, the $\alpha_+$ solution begins with $\alpha_+ = \infty$ at $T = -\infty$. It contracts to $\alpha_+ = \sqrt{\frac{3}{2}}$ at $T = \infty$.  The solution $\alpha_-$ has the opposite behavior, beginning at $\alpha_- = \sqrt{\frac{3}{2}}$ at $T = -\infty$ and expanding to an infinite size in the limit $T \to \infty$.

For $0 < A < \frac{3}{4}$, solutions reach extrema at
\be \alpha_{e_\pm} = \sqrt{\frac{3}{2} \pm \sqrt{3} \sqrt{\frac{3}{4}-A}} \;. \label{alphab} \ee
The solutions $\alpha_\pm$ both begin with $\alpha_\pm = \infty$ at $T = - \infty$, contract down to the same minimum size given by $\alpha_{e_+}$ and then expand to $\alpha_\pm = \infty$ in the limit $T \to \infty$.  There are also solutions that begin at $\alpha = 0$, expand to a maximum size given by $\alpha_{e_-}$ and then contract to $\alpha = 0$.

For $A > \frac{3}{4}$, the $\alpha_+$ solution begins with $\alpha_+ = \infty$ at $ T = -\infty$ and contracts to zero size at a finite value of $T$.  Then the $\alpha_-$ solution begins with $\alpha_- = 0$ at a finite value of $T$ and expands to $\alpha_- = \infty$ in the limit $T \to \infty$.

\end{document}